\documentclass[useAMS,referee,usenatbib]{biom}
\usepackage{amsmath}
\usepackage{graphicx,psfrag,epsf}
\usepackage{enumerate}
\usepackage{float}
\usepackage{url} 
\usepackage{xcolor}
\usepackage{amsmath}
\usepackage{amssymb}
\usepackage{amsfonts}
\usepackage{multirow}
\usepackage{subfig}
\usepackage{algorithm}
\usepackage[noend]{algpseudocode}

\newcommand{\argmax}{\normalfont\text{argmax}}
\newcommand{\argmin}{\normalfont\text{argmin}}
\newcommand{\lambdaF}{\lambda_{\normalfont\text{F}}}
\newcommand{\lambdaSp}{\lambda_{\normalfont\text{Sp}}}
\newcommand{\lambdaSim}{\lambda_{\normalfont\text{Sim}}}
\newtheorem{thm}{Theorem}[section]

\newtheorem{prop}{Proposition}[section]
\newcounter{assump}[section]
\newenvironment{assump}[1][]{\refstepcounter{assump}\par\medskip
	\noindent \textit{Assumption~\theassump. #1} \rmfamily}{\medskip}

\title[A Joint Fairness Model with Applications to Risk Predictions for Under-represented Populations]{A Joint Fairness Model with Applications to Risk Predictions for Under-represented Populations}

\author{Hyungrok Do$^{1}$, 
Shinjini Nandi$^{2}$,  
Preston Putzel$^{3}$,
Padhraic Smyth$^{3}$, and
Judy Zhong$^{1,*}$ \email{judy.zhong@nyulangone.org}
\\
$^{1}$Department of Population Health, NYU Grossman School of Medicine, New York, NY, 10016, USA \\
$^{2}$Department of Mathematical Sciences, Montana State University, Bozeman, MT, 59717, USA \\
$^{3}$Department of Computer Science, University of California, Irvine, CA, 92697, USA}

\begin{document}

\bibliographystyle{apalike}





\pagerange{\pageref{firstpage}--\pageref{lastpage}} 
\volume{(In Press)}
\pubyear{2022}
\artmonth{January}
\artmonth{Accepted in January}


\doi{-}


\label{firstpage}


\begin{abstract}
In data collection for predictive modeling, under-representation of certain groups, based on gender, race/ethnicity, or age, may yield less-accurate predictions for these groups. Recently, this issue of fairness in predictions has attracted significant attention, as data-driven models are increasingly utilized to perform crucial decision-making tasks. Existing methods to achieve fairness in the machine learning literature typically build a single prediction model in a manner that encourages fair prediction performance for all groups. These approaches have two major limitations: i) fairness is often achieved by compromising accuracy for some groups; ii) the underlying relationship between dependent and independent variables may not be the same across groups. We propose a Joint Fairness Model (JFM) approach for logistic regression models for binary outcomes that estimates group-specific classifiers using a joint modeling objective function that incorporates fairness criteria for prediction. We introduce an Accelerated Smoothing Proximal Gradient Algorithm to solve the convex objective function, and present the key asymptotic properties of the JFM estimates. Through simulations, we demonstrate the efficacy of the JFM in achieving good prediction performance and across-group parity, in comparison with the single fairness model, group-separate model, and group-ignorant model, especially when the minority group's sample size is small. Finally, we demonstrate the utility of the JFM method in a real-world example to obtain fair risk predictions for under-represented older patients diagnosed with coronavirus disease 2019 (COVID-19).
\end{abstract}

%

\begin{keywords}
algorithmic bias, algorithmic fairness, joint estimation, under-represented population \end{keywords}


\maketitle


%

\allowdisplaybreaks

\section{Introduction}
\subsection{Applied Context}

The issue of making fair predictions has attracted significant attention recently in machine learning as a critical issue in the application of data-driven models. Though machine learning models are increasingly utilized to perform crucial decision-making tasks, recent evidence reveals that many carefully designed algorithms can learn biases from the underlying data and exploit these inequities when making predictions. For example, large systematic biases in prediction performance have been detected for machine learning models in areas such as recidivism prediction relative to race \citep{angwin2016machine}, ranking of job candidates relative to gender \citep{DBLP:journals/corr/abs-1806-01059} and face recognition relative to both race and gender \citep{ryu2017inclusivefacenet,pmlr-v81-buolamwini18a}. 

There is an emerging recognition that such biases in data often leads to unfair performance from predictive models in healthcare for certain groups \citep{char2018implementing}, such as
women \citep{larrazabal2020gender}, ethnic and racial minorities \citep{seyyed2020chexclusion,doi:10.1146/annurev-biodatasci-092820-114757}, and individuals with public insurance \citep{seyyed2020chexclusion,doi:10.1146/annurev-biodatasci-092820-114757}.
Biased representations of different populations in biomedical studies, and the under-performance for under-represented populations, limit the potential benefits that can be achieved for these communities. In particular, when model-based predictions are used to prioritize patients for rationed services (e.g. organ transplantation, care management programs, or ICU services), under-performance for the under-represented patient populations will lead to unfair treatments for these patients \citep{paulus2020predictably}.

A particularly motivating example that we use in this paper is the mortality prediction for patients infected with coronavirus disease 2019 (COVID-19).
As of January 23 2021, COVID-19 has infected more than 96 million people globally, accounting for more than 2 million known deaths. Older patients are particularly vulnerable to severe outcomes and death due to COVID-19. 
The Centers for Disease Control and Prevention (CDC) reported that the fatality rate was 18.8\% for
patients older than 80 years whereas the overall fatality rate is estimated at  5\% or less for all patients \citep{CDCreport2}. 
This difference in survival highlights an urgent need for risk stratification of older patients with COVID-19 based on routine clinical assessments.
However, most COVID-19 studies have not been stratified by age groups \citep{TEHRANI2021415}. Thus, as an example, when applying a risk prediction equation generated from the general population to older patients with COVID-19, the model in \citet{TEHRANI2021415} predicts high-risk scores overall due to their older age, higher prevalence of comorbidities and more laboratory abnormalities. This resulted in insufficient and unfair risk stratification for these patients as not all older patients are at the same risk of death from COVID-19 \citep{TEHRANI2021415}.

\subsection{Existing Approaches}
Methods to address fairness in the machine learning literature typically begin with a formal probabilistic definition of fairness. In the context of risk prediction, predictive fairness at the group level means that a risk prediction model has performance characteristics (such as accuracy, ranking, or calibration) that are relatively independent of group membership. For example, if the false positive rate for a classification model is defined as $P(\hat{y}=1 | y=0)$, where $\hat{y}$ is the model’s prediction, then enforcing equality can be stated as requiring that these distributions be as close as possible between groups. Other definitions include demographic parity \citep{5360534}, equalized odds or equal opportunity \citep{hardt2016}, disparate treatment, and impact and mistreatment \citep{JMLR:v20:18-262,10.1145/3038912.3052660}. It is well-recognized that there is no unique optimal way to define fairness, leading to trade-offs between different approaches \citep{pmlr-v54-zafar17a}.

Given a fairness criterion, the second component of a fairness strategy requires an algorithmic approach, typically consisting of either 1) pre-processing the data by mapping the training data to a transformed space where the dependencies between sensitive attributes and class labels disappear \citep{KamiranCalders2012,dwork2018}; or 2) post-processing of a trained prediction model; for example, \citep{Kamishima2012,hardt2016} modify the probability of the decision being positive and negative predictions from an existing classifier to limit unfair discrimination; or 3) “in-process" methods, where fairness is accounted for during training of a model by adding a fairness constraint to the objective function. Examples of in-process methods include \citet{pmlr-v28-zemel13} who proposed to learn a fair representation of the data and classifier parameters by optimizing a non-convex function, and \citet{pmlr-v54-zafar17a} who defined a convex function as a measure of (un)fairness and suggested optimizing accuracy subject to the convex fairness constraints as well as their converse. 

A key feature of nearly all existing approaches is that a single set of classifier parameters is estimated using fairness criteria that encourage fair prediction performance across all groups. This approach has two main limitations: i) fairness is often achieved by compromising accuracy of some groups; ii) the underlying relationship between dependent and independent variables may  not  be  the  same  across groups, and the differences in predictive features may be of interest. 
In the example of predicting mortality risk for patients with COVID-19, while one would expect some features to have the same association with mortality for both older and younger patients, the associations between mortality and other features may be different between age groups. For instance, being overweight or obese (Body Mass Index [BMI] $>25 kg/m^2$) increases the risk for mortality associated with COVID-19, particularly among adults aged $<65$ years \citep{CDCreport2}
However, geriatric BMI guidelines are different from those for younger adults. For older adults, higher BMIs are often associated with greater energy stores and a better nutritional state overall, which is beneficial for patients' survival outcomes when serious infections develop. 

Estimating separate prediction models for each group does not leverage potential similarities between the groups. Moreover, estimating a single prediction model, even while using fairness criteria, will likely result in sub-optimal estimation or prediction performance for one group in order to achieve fair performance with a single set of parameters shared across groups.  
Outside the context of algorithmic fairness, \citet{danaher2014joint} proposed the joint graphical lasso method, a technique for jointly estimating multiple models corresponding to distinct but related conditions. Their approach is based upon a penalized log-likelihood approach, which penalizes the differences between parameter estimates across groups. 
Penalized log-likelihood approaches have often been used by other authors like \citet{Yuan2007}, \citet{Friedman2007} etc. for similar estimation purposes while minimizing the disparities in estimates across groups. In all such cases, however, prediction performance was not emphasized. 

In this paper, we propose the joint fairness model, a technique for jointly estimating multiple logistic regression models corresponding to distinct but related groups, in order to achieve fair prediction performance across groups. The model parameters are estimated by encouraging prediction fairness, while simultaneously ensuring high predictive accuracy irrespective of heterogeneity across  groups. The rest of this paper is organized as follows. In Section~\ref{sec:Problem}, we present the proposed joint fairness model.  Section~\ref{sec:Alg} describes the algorithm to find its optimal solution. In Section~\ref{sec:Theory}, we discuss asymptotic consistency of the estimators. We illustrate the performance of our proposed approach in simulation studies in Section~\ref{sec:Simu}; and apply our approach to the motivating example of predicting COVID-19 mortality outcomes for patients of different age groups in Section~\ref{sec:app}. Finally, we summarize and discuss our findings in Section~\ref{sec:con}.

\section{Problem Formulation}\label{sec:Problem}
For binary outcomes, consider $K$ groups of datasets $S_{k}=\{(\textbf{X}_{ki}, y_{ki})\in\mathbb{R}^{p} \times \{0,1\}:i=1,\cdots,n_{k}\}$ with $K \geq 2$ representing group membership. Throughout the paper, group memberships are known and observed. Assuming that the $n=\sum_{k=1}^{K}{n_k}$ observations are independently distributed, then $y_{ki} \sim \text{Bernoulli}(\text{p}_{ki})$, and $\hat{y}_{ki}:\mathbb{R}^{p} \rightarrow \{0,1\}$ is the predicted value based on predictor features $\mathbf{X}_{ki}$. 
We focus on the development of a fair prediction approach for the widely-used logistic regression model.  The log-likelihood of the logistic model for the data from all groups takes the form 
\begin{equation}\label{eqn:over}
\sum_{k=1}^{K}\ell(\boldsymbol{\beta}_k;\mathbf{X}_{k}, \mathbf{y}_{k})=\sum_{k=1}^{K}\sum_{i=1}^{n_k}\left(y_{ki}\mathbf{X}_{ki}\boldsymbol{\beta}_{k} - \log \left( 1 + \exp \left(\mathbf{X}_{ki}\boldsymbol{\beta}_{k} \right)\right)\right).
\end{equation}
Define $\boldsymbol{\beta}= (\boldsymbol{\beta}_{1}, \cdots, \boldsymbol{\beta}_{K}) \in \mathbb{R}^{pK}$. Maximizing the likelihood function in \eqref{eqn:over} with respect to $\boldsymbol{\beta}_k$ in each group separately yields the maximum likelihood estimates of parameters $\hat{\boldsymbol{\beta}}_k$ for each group $k$, thus making separate predictions $\hat{y}_k$ per group. If we ignore group memberships, $\hat{\boldsymbol{\beta}}$ can be estimated by maximizing the likelihood function in equation \eqref{eqn:over} setting all $\boldsymbol{\beta}_k$ equal to a single global parameter vector $\hat{\boldsymbol{\beta}}$ and making predictions $\hat{y}$ per individual (irrespective of group) using that global parameter vector.

If the $K$ datasets correspond to observations collected from $K$ distinct but related groups, then one might wish to borrow strength across the $K$ groups to estimate $\boldsymbol{\beta}$ and predict $\hat{y}$, rather than estimating parameters $\boldsymbol{\beta}_k$ for each group separately, or estimating a single set of $\boldsymbol{\beta}_k$ for all $k$ which could lead to heterogeneous prediction performance across the groups. 
Therefore, instead of estimating $\boldsymbol{\beta}$ by maximizing the likelihood in equation \eqref{eqn:over}, we consider a penalized log-likelihood approach and jointly estimate $\boldsymbol{\beta}$ by maximizing an objective function of $\sum_{k=1}^{K}\ell(\boldsymbol{\beta}_k; \mathbf{X}_{k}, \mathbf{y}_{k})$ in equation \eqref{eqn:over} subject to constraints on (i) fairness, $\mathcal{P}_{\text{F}}(\boldsymbol{\beta}; \mathbf{X}, \mathbf{y}, \lambdaF)$ (ii) parameter similarity, $\mathcal{P}_{\text{Sim}}(\boldsymbol{\beta}; \lambdaSim)$, and (iii) parameter sparsity, $\mathcal{P}_{\text{Sp}}(\boldsymbol{\beta}; \lambdaSp)$. 
\begin{equation}\label{eqn:formulation2}
\underset{\boldsymbol{\beta} \in \mathbb{R}^{pK}}{\text{minimize}} ~ F(\boldsymbol{\beta}) = -\sum_{k=1}^{K}\frac{1}{n_{k}}\ell(\boldsymbol{\beta}_{k}; \mathbf{X}_{k}, \mathbf{y}_{k})  
+ \mathcal{P}_{\text{F}}(\boldsymbol{\beta}; \mathbf{X}, \mathbf{y}, \lambdaF)
+ \mathcal{P}_{\text{Sim}}(\boldsymbol{\beta}; \lambdaSim)
+ \mathcal{P}_{\text{Sp}}(\boldsymbol{\beta}; \lambdaSp)
.
\end{equation}

We propose a fairness penalty function $\mathcal{P}_{\text{F}}(\boldsymbol{\beta}; \mathbf{X}, \mathbf{y}, \lambdaF)$ that encourages each group to have similar predictive performance. In this work  we use equalized odds \citep{hardt2016} which encourages each group to have similar false positive rates (FPRs) and false negative rates (FNRs).  
Thus, we want to minimize the absolute difference between $\text{FPR}_{j}$ and $\text{FPR}_{k}$ $| P(\hat{y}_j=1|y_j=0)$ $-$ $P(\hat{y}_k=1|y_k=0)|$, and that between $\text{FNR}_{j}$ and $\text{FNR}_{k}$: $| P(\hat{y}_j=0|y_j=1)$ $-$ $P(\hat{y}_k=0|y_k=1) |$. 
Under the logistic regression model, the absolute differences of FPRs and FNRs are nonconvex due to the nonconvexity of the sigmoid function. We will instead minimize the absolute difference of the expected linear components of the two groups $\Big|\mathbb{E}\big[\mathbf{X}_j\boldsymbol{\beta}_{j}\big|{y}_j=0 \big] - \mathbb{E}\big[\mathbf{X}_k\boldsymbol{\beta}_{k}\big|{y}_k=0 \big]\Big|$.
The proposition below shows that the absolute difference of the expected probabilities is upper-bounded by the absolute difference of the expected linear components. Thus, minimizing group differences in expected linear components guarantees that group difference of false predictions is minimized.   
\begin{prop}
For any $1 \leq j, k \leq K, j \neq k$, and $y \in \{0,1\}$ the following inequality holds:
    \begin{align}\nonumber
    \Bigg|\mathbb{E}\Bigg[\frac{1}{1+\exp(-\mathbf{X}_j\boldsymbol{\beta}_{j})}\Bigg|y_j=y \Bigg] - & \mathbb{E}\Bigg[\frac{1}{1+\exp(-\mathbf{X}_k\boldsymbol{\beta}_{k})}\Bigg|y_k=y \Bigg]\Bigg| \\
    \nonumber & \leq \frac{1}{4}\left|\mathbb{E}\left[{\mathbf{X}_j\boldsymbol{\beta}_{j}} \Bigg|y_j=y\right] - \mathbb{E}\left[{\mathbf{X}_k\boldsymbol{\beta}_{k}} \Bigg|y_k=y\right]\right|.
\end{align}
\end{prop}
Proof: See Web Appendix 4.

Note that the empirical estimate of the expectation is
\begin{equation}\nonumber
    \mathbb{E}\Big[\mathbf{X}_k\boldsymbol{\beta}_{k}\Big|y_k=y\Big] = \frac{1}{|S_{ky}|} \sum_{i \in S_{ky}} \mathbf{X}_{i}\boldsymbol{\beta}_{k},
\end{equation}
where $S_{ky}=\{i:y_{ki} = y\}$ is a subgroup of subjects with a true response value $y$ in group $k$,  with $y \in \{0,1\}$. Thus, our fairness penalty, that bridges the between-group gaps in the linear components of FPR$_{k}$ and FNR$_{k}$, is defined as: 
\begin{align}\label{eqn:fair}
    \mathcal{P}_{\text{F}}(\boldsymbol{\beta}; &\mathbf{X}, \mathbf{y}, \lambda_{\text{F}}) 
    = \mathcal{P}_{\text{FPR}}(\boldsymbol{\beta}; \mathbf{X}, \mathbf{y}, \lambda_{\text{F}}) + \mathcal{P}_{\text{FNR}}(\boldsymbol{\beta}; \mathbf{X}, \mathbf{y}, \lambda_{\text{F}}) \\ \nonumber
    &= \lambda_{\text{F}}\sum_{j < k}\left|\frac{1}{|S_{j0}|}\sum_{i\in S_{j0}} \mathbf{X}_{i}\boldsymbol{\beta}_{j} - \frac{1}{|S_{k0}|}\sum_{i \in S_{k0}} \mathbf{X}_{i}\boldsymbol{\beta}_{k} \right|
    + \lambda_{\text{F}}\sum_{j < k}\left|\frac{1}{|S_{j1}|}\sum_{i\in S_{j1}} \mathbf{X}_{i}\boldsymbol{\beta}_{j} - \frac{1}{|S_{k1}|}\sum_{i \in S_{k1}} \mathbf{X}_{i}\boldsymbol{\beta}_{k} \right|
\end{align} 
where the summation $\sum_{j<k}$ represents $\sum_{k=2}^{K}\sum_{j=1}^{k-1}$ for notational simplicity. 

The similarity penalty $\mathcal{P}_{\text{Sim}}(\boldsymbol{\beta};\lambda_{\text{Sim}})$ is chosen to encourage similarity across the $K$ parameter estimates. Here we use the generalized fused lasso penalty \citep{doi:10.1198/jcgs.2010.09208,danaher2014joint,10.1093/biostatistics/kxy035} defined as 
\begin{align}\label{eqn:sim}
	\mathcal{P}_{\text{Sim}}(\boldsymbol{\beta}; \lambdaSim) = \lambda_{\text{Sim}}\sum_{{j < k}}\|\boldsymbol{\beta}_{j} - \boldsymbol{\beta}_{k}\|_{1},
\end{align}
herein referred to as the fusion-similarity penalty. The sparsity penalty $\mathcal{P}_{\text{Sp}}(\boldsymbol{\beta})$ is chosen to encourage sparse estimates and to avoid ill-defined maximum likelihood estimates when sample sizes $n_k < p$.  
\begin{equation}\label{eqn:lasso}
    \mathcal{P}_{\text{Sp}}(\boldsymbol{\beta};\lambdaSp) = \sum_{k=1}^{K}\lambda_{\text{Sp}_k}\|\boldsymbol{\beta}_{k}\|_{1}.
\end{equation}
In the three penalty functions, $\lambdaF$, $\lambdaSim$, and $\lambdaSp$ are nonnegative hyperparameters. 
Here $\mathcal{P}_{\text{Sp}}(\boldsymbol{\beta};\lambda_{\text{Sp}})$, $\mathcal{P}_{\text{F}}(\boldsymbol{\beta}; \mathbf{X}, \mathbf{y}, \lambda_{\text{F}})$, and  $\mathcal{P}_{\text{Sim}}(\boldsymbol{\beta};\lambda_{\text{Sim}})$ are convex penalty functions, so that the objective function in equation \eqref{eqn:formulation2} is convex in $\boldsymbol{\beta}$. The proposed model jointly estimates $\boldsymbol{\beta}$ to achieve fair performance across groups, herein referred to as the joint fairness model (JFM). In contrast, the dominant approach for fair predictions in the current literature is to estimate a single set of $\boldsymbol{\beta}$ parameters with constraints on the quality of performance metrics across groups. 
In the context of logistic regression, \citet{bechavod2017penalizing} proposed a Single Fairness Model (SFM) to minimize the following objective function. 
\begin{align}\nonumber
\underset{\boldsymbol{\beta} \in \mathbb{R}^{p}}{\text{minimize}} ~ -\sum_{k=1}^{K}\ell(\boldsymbol{\beta};\mathbf{X}_{k}, \mathbf{y}_{k}) +
\lambda_{\text{F}} \sum_{j < k} \sum_{y \in \{0,1\}} \left|\frac{1}{|S_{jy}|}\sum_{i \in S_{jy}} \mathbf{X}_{i}\boldsymbol{\beta} - \frac{1}{|S_{ky}|} \sum_{i\in S_{ky}} \mathbf{X}_{i}\boldsymbol{\beta} \right|  + \lambda_{\text{Sp}}\|\boldsymbol{\beta}\|_{1}.
\end{align} 
In contrast to the proposed JFM, it does not allow for group-specific different $\boldsymbol{\beta}_k$s, leading to two potential limitations: i) fairness is often achieved by compromising the likelihood ($\ell(\boldsymbol{\beta};\mathbf{X}_{k}, \mathbf{y}_{k})$) of some groups; ii) inflexible model mis-specification when the $\boldsymbol{\beta}_k$s are different.

The proposed JFM objective function improves prediction parity through three components. First, it considers a weighted group total likelihood to upweight the groups with smaller sample sizes. Second, $\mathcal{P}_{\text{F}}(\boldsymbol{\beta}; \mathbf{X}, \mathbf{y}, \lambda_{\text{F}})$ encourages estimates that achieve fair performance across groups. Third, the similarity penalty term improves estimation and prediction efficiencies when multiple subgroups are related but not identical \citep{doi:10.1198/jcgs.2010.09208,danaher2014joint}. Computationally, we found that multiple different combinations of $\hat{\boldsymbol{\beta}}_k$ often result in very similar objective values; thus, the similarity term can help optimization algorithms for the JFM converge to one of the multiple combinations by favoring similar values of $\hat{\boldsymbol{\beta}}_{k}$. Other formats of the similarity penalty could be used in our proposed JFM framework. For example, the group lasso penalty \citep{yuan2006model} has been shown to encourage similar sparsity patterns across groups \citep{obozinski2010joint,danaher2014joint}, while the fused lasso term is more aggressive in encouraging similar $\hat{\boldsymbol{\beta}}_k$ estimates. 
The penalty functions in \eqref{eqn:fair}, \eqref{eqn:sim}, and \eqref{eqn:lasso} are based on the $L_{1}$ norm. 
They can be flexibly adapted to an $L_{2}$ penalization or to a combination of $L_{1}$ and $L_{2}$ penalizations. 
The differences between $L_{1}$ and $L_{2}$ penalties have been well discussed in the literature \citep{https://doi.org/10.1111/j.2517-6161.1996.tb02080.x,10.2307/3647580}. For the fairness penalty, \citet{bechavod2017penalizing} showed that there are no significant differences in the empirical predictive performances between the $L_{1}$ and $L_{2}$ fairness penalty forms other than the $L_{2}$ form of the similarity penalty penalizes large differences more aggressively so that models have less chance to obtain group-specific estimates.
\section{Accelerated Smoothing Proximal Gradient Algorithm for JFM}\label{sec:Alg}
In this section, we introduce an accelerated smoothing proximal gradient (ASPG) algorithm \citep{chen2012smoothing} to solve Problem \eqref{eqn:formulation2} for the JFM. The objective function of \eqref{eqn:formulation2} is convex in $\boldsymbol{\beta}$ so that a global optimal solution can be attained. However, conventional proximal gradient-based or coordinate descent approaches (generally used for lasso-like methods) cannot be directly applied to solve Problem \eqref{eqn:formulation2} because there is no closed-form solution for the proximal operator associated with $\mathcal{P}_{\text{FPR}}$ and $\mathcal{P}_{\text{FNR}}$. 

\subsection{Nesterov smooth approximation}
To overcome the difficulty originating from the non-differentiability of the fairness and similarity penalties, we decouple the terms into a linear combination of the decision variables via the dual norm and then apply the Nesterov smoothing approximation \citep{nesterov2005smooth}.
We start with matrix representations of the fairness penalty terms $\mathcal{P}_{\text{FPR}}(\boldsymbol{\beta};\mathbf{X}, \mathbf{y}, \lambda_{\text{F}}) = \lambda_{\text{F}}\left\| \mathbf{D}_{0} \boldsymbol{\beta} \right\|_{1}$ and $\mathcal{P}_{\text{FNR}}(\boldsymbol{\beta};\mathbf{X}, \mathbf{y}, \lambda_{\text{F}}) = \lambda_{\text{F}}\left\| \mathbf{D}_{1} \boldsymbol{\beta} \right\|_{1}$, where $\mathbf{D}_{y} \in \mathbb{R}^{K(K-1)/2 \times pK}$ is defined as below. Similarly, we use the matrix representation of the similarity penalty $\mathcal{P}_{\text{Sim}}(\boldsymbol{\beta};\lambda_{\text{Sim}}) = \lambda_{\text{Sim}}\left\| \mathbf{F} \boldsymbol{\beta} \right\|_{1}$ with $\mathbf{F}$ defined as below.
\begin{equation}\nonumber
    {\scriptsize \mathbf{D}_{y} = \left(\begin{array}{ccccc}
    \bar{\mathbf{X}}_{1y} & -\bar{\mathbf{X}}_{2y} & \mathbf{0} & \cdots & \mathbf{0} \\
    &&\vdots&&\\
    \mathbf{0} & \bar{\mathbf{X}}_{2y} & -\bar{\mathbf{X}}_{3y} & \cdots & \mathbf{0} \\
    &&\vdots&&
    \end{array}\right) ~~~
    \mathbf{F}=\left(\begin{array}{ccccc}
    \mathbf{I}_{p} & -\mathbf{I}_{p} & \mathbf{0} & \cdots & \mathbf{0} \\
    &&\vdots&&\\
    \mathbf{0} & \mathbf{I}_{p} & -\mathbf{I}_{p} & \cdots & \mathbf{0} \\
    &&\vdots&&
    \end{array}\right)}
\end{equation}

Here, $\bar{\mathbf{X}}_{ky} = \frac{1}{|S_{ky}|} \sum_{i \in S_{ky}} \mathbf{X}_{i}$ is the average predictor vector for group $k$ with true outcome $y$, and $\mathbf{I}_{p}$ is the $p$-dimensional identity matrix.
The matrix form of the fairness penalty term and the similarity penalty term is therefore defined as:
\begin{equation}\nonumber
    {\mathcal{P}_{\text{F}}(\boldsymbol{\beta};\mathbf{X},\mathbf{y},\lambda_{\text{F}}) +  \mathcal{P}_{\text{Sim}}(\boldsymbol{\beta};\lambda_{\text{Sim}})} = \left\|
    \left(\begin{array}{c}
         \lambda_{\text{F}}\mathbf{D}_{0} \\
         \lambda_{\text{F}}\mathbf{D}_{1} \\
         \lambda_{\text{Sim}}\mathbf{F} 
    \end{array}\right) \boldsymbol{\beta} ~
    \right\|_{1} = {\|\mathbf{D}_{\lambda_{\text{F}},\lambda_{\text{Sim}}}\boldsymbol{\beta}\|_{1}}.
\end{equation}
Thus, the objective function \eqref{eqn:formulation2} can be written in matrix form:
\begin{equation}\label{matrxiobjective}
    \underset{\boldsymbol{\beta}}{\text{minimize}} ~ -\sum_{k=1}^{K}\frac{1}{n_{k}}\ell(\boldsymbol{\beta}_{k}; \mathbf{X}_{k}, \mathbf{y}_{k}) + {\|\mathbf{D}_{\lambda_{\text{F}},\lambda_{\text{Sim}}}\boldsymbol{\beta}\|_{1}} +  \sum_{k=1}^{K} \lambda_{\text{Sp}_{k}} \|\boldsymbol{\beta}_{k}\|_{1},
\end{equation}
where the associated proximal operator of ${\|\mathbf{D}_{\lambda_{\text{F}},\lambda_{\text{Sim}}}\boldsymbol{\beta}\|_{1}}$ does not have a closed form solution.
We apply the Nesterov smooth approximation to approximate ${\|\mathbf{D}_{\lambda_{\text{F}},\lambda_{\text{Sim}}}\boldsymbol{\beta}\|_{1}}$ by a smooth function 
\begin{align}\label{eq7}
    f_{\mu}(\boldsymbol{\beta};\lambda_{\text{F}}, \lambda_{\text{Sim}}) = \sup\left\{\boldsymbol{\alpha}^{T}{\mathbf{D}_{\lambda_{\text{F}},\lambda_{\text{Sim}}}\boldsymbol{\beta}} - \frac{\mu}{2}\|\boldsymbol{\alpha}\|_{2}^{2}:\|\boldsymbol{\alpha}\|_{\infty} \leq 1\right\}.
\end{align}

Proposition A.1 in Web Appendix 1 provides the maximum gap between $\|\mathbf{D}_{\lambda_{\text{F}}, \lambda_{\text{Sim}}}\boldsymbol{\beta}\|_{1}$ and its Nesterov approximation $f_{\mu}(\boldsymbol{\beta};\lambda_{\text{F}}, \lambda_{\text{Sim}})$. 

As demonstrated in Proposition A.2 in Web Appendix 1, for any $\mu > 0$, $f_{\mu}(\boldsymbol{\beta};\lambdaF, \lambdaSim)$ is smooth and convex with respect to $\boldsymbol{\beta}$, whose gradient takes the following form:
\begin{equation}\label{eq8}
    \nabla f_{\mu}(\boldsymbol{\beta};\lambdaF, \lambdaSim) = {\mathbf{D}_{\lambda_{\normalfont{\text{F}}},\lambda_{\normalfont{\text{Sim}}}}^{T}}\boldsymbol{\alpha}^{*},
\end{equation}
where $\boldsymbol{\alpha}^{*}=\argmax \left\{\boldsymbol{\alpha}^{T}{\mathbf{D}_{\lambda_{\normalfont{\text{F}}},\lambda_{\normalfont{\text{Sim}}}}}\boldsymbol{\beta} - \frac{\mu}{2}\|\boldsymbol{\alpha}\|_{2}^{2}:\|\boldsymbol{\alpha}\|_{\infty} \leq 1\right\}$. Moreover, the gradient is Lipschitz continuous with a Lipschitz constant $L_{\mu} = \mu^{-1}\|{\mathbf{D}_{\lambda_{\normalfont{\text{F}}},\lambda_{\normalfont{\text{Sim}}}}}\|_{2}^{2}$, where $\|\cdot\|_{2}$ denotes the matrix spectral norm (which is equivalent to the largest singular value of the matrix). We can show further that $\boldsymbol{\alpha}^{*}$ can be calculated as $\boldsymbol{\alpha}^{*} = \mathcal{S}_{\infty}\left( \mu^{-1}{\mathbf{D}_{\lambda_{\normalfont{\text{F}}},\lambda_{\normalfont{\text{Sim}}}}}\boldsymbol{\beta} \right)$,
where $\mathcal{S}_{\infty}(\cdot)$ is the projection onto the unit $L_{\infty}$ ball such that $[\mathcal{S}_{\infty}(\boldsymbol{x})]_{i} = x_{i} \mathbb{I}_{[-1,1]}(x_{i}) + \mathbb{I}_{(1,\infty)}(x_{i}) - \mathbb{I}_{(-\infty,-1)}(x_{i})$, where $\mathbb{I}$ is the indicator function.
Details are provided in Web Appendix 1. 

\subsection{Accelerated Smoothing Proximal Gradient Algorithm}
With ${\|\mathbf{D}_{\lambda_{\normalfont{\text{F}}},\lambda_{\normalfont{\text{Sim}}}}\boldsymbol{\beta}\|_{1}}$ substituted by the Nesterov smooth approximation $f_{\mu}(\boldsymbol{\beta};\lambdaF, \lambdaSim)$, problem \eqref{matrxiobjective} becomes
\begin{equation}\label{eqn:approx}
    \underset{\boldsymbol{\beta}}{\text{minimize}} ~  \tilde{F}(\boldsymbol{\beta}) = -\sum_{k=1}^{K} \frac{1}{n_{k}}\ell(\boldsymbol{\beta}_{k}; \mathbf{X}_{k}, \mathbf{y}_{k}) + f_{\mu}(\boldsymbol{\beta};\lambdaF, \lambdaSim) + \sum_{k=1}^{K} \lambda_{\text{\normalfont{Sp}}_{k}} \|\boldsymbol{\beta}_{k}\|_{1} ,
\end{equation}
whose first two terms are convex smooth functions. Although the sparsity penalty term $\sum_{k}\lambda_{\text{\normalfont{Sp}}_{k}} \|\boldsymbol{\beta}_{k}\|_{1}$ is non-differentiable, it can be managed through the proximal gradient method using the soft-thresholding operator $\mathcal{S}$ with a closed form solution \citep{friedman2007pathwise}.

\begin{center} [Algorithm \ref{pseudocode:JointFair} about here.] \end{center}
Algorithm \ref{pseudocode:JointFair} presents the proposed ASPG algorithm, starting from parameter initialization, to gradient descent iterations with proximal and momentum steps, until convergence. Although Algorithm 1 minimizes the Nesterov smooth approximation, Theorem \ref{theorem:conv} proves that the solution can reach arbitrarily close to the global optimum of Problem \eqref{eqn:formulation2}.
\begin{thm}\label{theorem:conv}
    Let $\{\boldsymbol{\beta}^{(t)}:t=1,2,\cdots\}$ be a sequence generated by Algorithm \ref{pseudocode:JointFair}. Then for any $t \geq 1$ and desired $\delta>0$
    \begin{equation}\nonumber
        F(\boldsymbol{\beta}^{(t)}) - F(\boldsymbol{\beta}^{**}) \leq \delta + \frac{2L\|\boldsymbol{\beta}^{(0)}-\boldsymbol{\beta}^{*}\|^{2}_{2}}{t^{2}},
    \end{equation}
where $\boldsymbol{\beta}^{*}$ and $\boldsymbol{\beta}^{**}$ are global minimizers of Problem \eqref{eqn:approx} and Problem \eqref{eqn:formulation2}, respectively and $L$ is the Lipschitz constant of the function in \eqref{eq7}.
\end{thm}
Proof: See Web Appendix 4.

Based on Theorem \ref{theorem:conv}, the rate of the convergence of Algorithm 1 is $\mathcal{O}\left(\sqrt{\frac{pK}{\delta(\varepsilon - \delta)}}\right)$, given a desired accuracy $\varepsilon > 0$. The complexity of a single iteration of Algorithm \ref{pseudocode:JointFair} is $\mathcal{O}((n+K^{2})pK)$. For additional details see Propositions A.4 and A.5 in Web Appendix 1. The proposed JFM Algorithm 1 is for the fusion-similarity term. The algorithm can be readily extended to include a group-similarity term as presented in Web Appendix 2.

\section{Asymptotic properties of the JFM estimates}\label{sec:Theory}
We now present  asymptotic results for the JFM parameter estimates $\hat{\boldsymbol{\beta}}$ obtained by solving Problem \eqref{eqn:formulation2}. We assume $p$ remains constant and $n$ increases to infinity.

Consider the following assumptions:
\begin{assump}\label{assumption1}
$\mathcal{I}(\boldsymbol{\beta}_k)/n_k \rightarrow \mathbf{C}_k$, where $\mathbf{C}_k$ is a positive definite $p\times p$ matrix, for $k=1,\cdots,K$, where $\mathcal{I}(\boldsymbol{\beta}_k)$ is the information matrix of size $p\times p$. For simplicity, we assume there are no intercept terms in $\boldsymbol{\beta}_k$.
\end{assump}

\begin{assump}\label{assumption2}
As $n =\underset{k=1,\cdots,K}{\min}~n_k \rightarrow \infty$, $\underset{\hat{\boldsymbol{\beta}}_k}{\max}\left\|\left(\mathcal{I}(\hat{\boldsymbol{\beta}}_k) ^{-\frac{1}{2}}\right) \mathcal{I}(\boldsymbol{\beta}_k)\left(\mathcal{I}(\hat{\boldsymbol{\beta}}_k)^{-\frac{1}{2}}\right)^T - \mathbf{I}_{p} \right\|_{2} \rightarrow 0$,
\end{assump}
where $\mathcal{I}(\hat{\boldsymbol{\beta}}_k)$ is the empirical information matrix, and $\mathbf{I}_p$ is a $p \times p$ identity matrix.

The following theorem proves $\sqrt{n}$-consistency for the estimators, complying with the fairness and similarity constraints between the two groups as well as the sparsity constraint. We note that the theorem holds even if the sample size of one group increases faster than the other group's.

\begin{thm}\label{thm:asymptotic}
Let $\hat{\boldsymbol{\beta}}_k$ for $k=1, \cdots, K$, minimize the loss function (\ref{eqn:formulation2}). If $\lambdaF^{(n)}/\sqrt{n} \rightarrow \lambdaF^{(0)} \ge 0$, 
$\lambdaSim^{(n)}/\sqrt{n} \rightarrow \lambdaSim^{(0)} \ge 0$, and
$\lambdaSp^{(n)}/\sqrt{n} \rightarrow \lambdaSp^{(0)} \ge 0$, then under the assumptions \ref{assumption1} and \ref{assumption2}
\begin{align}\label{eqn:consistency}
	\sqrt{n} \left( \hat{\boldsymbol{\beta}}_k - \boldsymbol{\beta}_k\right) \overset{d}{\rightarrow} \mathbf{\hat{u}}_k,
\end{align}
where $(\mathbf{\hat{u}}_{1}, \cdots, \mathbf{\hat{u}}_{K})=\argmin~\mathcal{V}(\mathbf{u}_{1}, \cdots, \mathbf{u}_{K})$,
\begin{align*}
	\mathcal{V}(\mathbf{u}_{1}, \cdots, \mathbf{u}_{K})  &= 
    \sum_{k=1}^{K}\mathbf{u}_{k}^{T}\mathbf{W}_{k} + \frac{1}{2} \sum_{k=1}^{K} \mathbf{u}_{k}^{T}\mathbf{C}_{k}\mathbf{u}_{k} \\& 
	+ \lambdaF^{(0)}\sum_{j < k} \sum_{y \in \{0,1\}}\mathcal{T}(\bar{\mathbf{X}}_{jy}\mathbf{u}_{j}-\bar{\mathbf{X}}_{ky}\mathbf{u}_{k},\bar{\mathbf{X}}_{jy}\boldsymbol{\beta}_{j}-\bar{\mathbf{X}}_{ky}\boldsymbol{\beta}_{k}) \\&
    + \lambdaSim^{(0)}\sum_{j<k}\sum_{l=1}^{p} \mathcal{T}(u_{jl} - u_{kl}, \beta_{jl} - \beta_{kl})
    + \lambdaSp^{(0)}\sum_{k=1}^{K}\sum^p_{l=1}\mathcal{T}(u_{kl},\beta_{kl}).
\end{align*}
Here $\mathcal{T}(u,\beta) = u \cdot \text{\normalfont sign}(\beta) \cdot \mathbb{I}(\beta \neq 0) + |u| \cdot \mathbb{I}(\beta = 0)$, and $\mathbf{W}_k\sim \mathcal{N}_{p}(\mathbf{0}, \mathbf{C}_k)$, where $\mathbf{C}_{k}=\underset{n_{k} \rightarrow \infty}{\lim}\frac{1}{n_{k}}\sum_{i=1}^{n_{k}}\mathbf{X}_{ki}\mathbf{X}_{ki}^{T}$, and  $\frac{1}{|S_{ky}|}\sum\limits_{i \in S_{ky}}\mathbf{X}_i = \bar{\mathbf{X}}_{ky}$, for $k = 1,\cdots,K$ and $y=0,1$.
\end{thm}
Proof: See Web Appendix 4.

\section{Simulation Study}\label{sec:Simu}

We performed a series of simulations to evaluate the proposed JFM, and compared it with the approaches of a group-separate individual logistic regression model, a group-ignorant vanilla logistic regression model,
and the SFM method \citep{bechavod2017penalizing} implemented using an SFM-ASPG algorithm (see Web Appendix 3). When applying the group-separate model, regression coefficients were estimated for each group separately with an $L_{1}$ penalty. The group-ignorant model estimates a single logistic regression with group membership as an additional covariate with an $L_{1}$ penalty. 

\subsection{Simulation Setup}
We consider a two-group problem ($K=2$) for simplicity, with group 1 as the over-represented group and group 2 as the under-represented group with respect to the sample sizes. 
The training samples were simulated as follows. The predictor matrix $\mathbf{X}_{k}$ was independently generated from a standard normal distribution. The binary outcome $y_{ki}$ was then simulated from $\text{Bernoulli}(\pi_{i}(\textbf{x}_{ki}))$, where $\pi_{i}(\textbf{x}_{ki}) = \frac{ \exp(\mathbf{x}_{ki} \boldsymbol{\beta}_{k}) }{ 1 + \exp(\mathbf{x}_{ki} \boldsymbol{\beta}_{k})}$. 
Out of the total number of features, 40\% in each group had non-zero coefficients ($\beta$'s). The non-zero coefficients were each set to the value 3. The simulations were conducted under  three scenarios to investigate performances at various levels of shared parameters, sample sizes and dimensionalities.
\begin{itemize}
\item \textbf{Scenario 1 (Difference in True Model):} The proportion of shared features between the two groups ranged from $0\%$ to $100\%$ of features with non-zero coefficients. The intercepts were selected so that the baseline event prevalences were at 30\% and 50\% for the under- and over-represented groups The sample sizes were set at $500$ and $200$ for group 1 and 2 respectively. The number of features was set to $p = 100$.
Note that when small proportions of features are shared between groups, the fairness and similarity penalty terms are mis-specified given that  $\mathbf{X}\boldsymbol{\beta}_k$ and $\boldsymbol{\beta}_k$ are different between groups. As a result,  performances in these settings allow us to investigate the robustness of the JFM approach to mis-specification of the fairness and similarity penalty terms.
    
\item \textbf{Scenario 2 (Difference in Sample Size):} The sample size of the under-represented group (group 2) ranged from $50$ to $300$ while the sample size of group 1 was fixed at $500$. The number of features was set to $p = 100$. Half of the features with non-zero coefficients were shared between the groups.
\item \textbf{Scenario 3 (High Dimensionality):} The number of features $p$ ranged from 50 to 2,000. Sample sizes were $500$ and $200$ for group 1 and 2 respectively. For each value of $p$, 40 features had non-zero coefficients, with half of the non-zero features being shared.
\end{itemize}
We evaluated the methods on independent test  datasets with large sample sizes ($n = 1,000$ for both groups) under the same simulation setups. The Area under the Receiver Operating Characteristic curve (AUC) was used to assess the predictive ability of each model. Prediction unfairness was assessed by the group difference in AUCs. In addition, Mean Squared Errors (MSEs) for all $\hat{\boldsymbol{\beta}}$s were examined to assess parameter estimation performance, and TPRs of the associated features and TNRs of the non-associated features were used to assess variable selection performance. Medians and interquartile ranges (IQRs) of the assessment metrics were generated from $20$ replicates for each experiment. Predictive performances and their unfairness in terms of FPR and FNR were calculated with a   predicted probability cutoff of $0.5$, as presented in Web Appendix 8. We present additional simulation scenarios in Web Appendix 9.

\subsection{Choice of the Hyperparameters}
The group-ignorant model, group-separate model, SFM, and JFM contain 1, $K$, $2$, and $K+2$ hyperparameters respectively. For every method, five-fold cross-validation on the training dataset was used to determine the hyperparameters. For the vanilla models (group-separate and group-ignorant), the lasso penalty term was selected by optimizing cross-validation AUCs. 
For the fairness-aware models (SFM and JFM), we investigated and compared a series of evaluation metrics for selecting  hyperparameters using cross-validation. The metrics include overall AUCs and Brier scores (defined as $\sum(\hat{p}_{ki}-y_{ki})^2$) on all samples (ignoring group memberships), group average of AUCs or Brier scores, and the group average of AUCs or Brier scores subtracting their disparities. Web Appendix 7, Web Figures 2, 3, and 4 demonstrate  the empirical performance  of the hyperparameters selected by the various strategies in the test datasets for the simulation scenarios. In summary, we find that the hyperparameters optimizing cross-validated group-average metrics  showed better performance  than sample-average or average metrics subtracting disparities. The hyperparameters optimizing AUCs in general generated the best AUC performances, while the hyperparameters optimizing Brier scores generated the best MSEs from the perspective of parameter estimation. We find that both are better empirically than threshold-based metrics such as classification accuracy. Lastly, the group average calculated by the harmonic mean is more robust than the arithmetic mean when the group sample sizes are unbalanced.

\subsection{Simulation Results}
\begin{figure}[!t]
    \centering
    \subfloat[AUC of the Under-represented Group]{\includegraphics[width=0.4\linewidth, keepaspectratio]{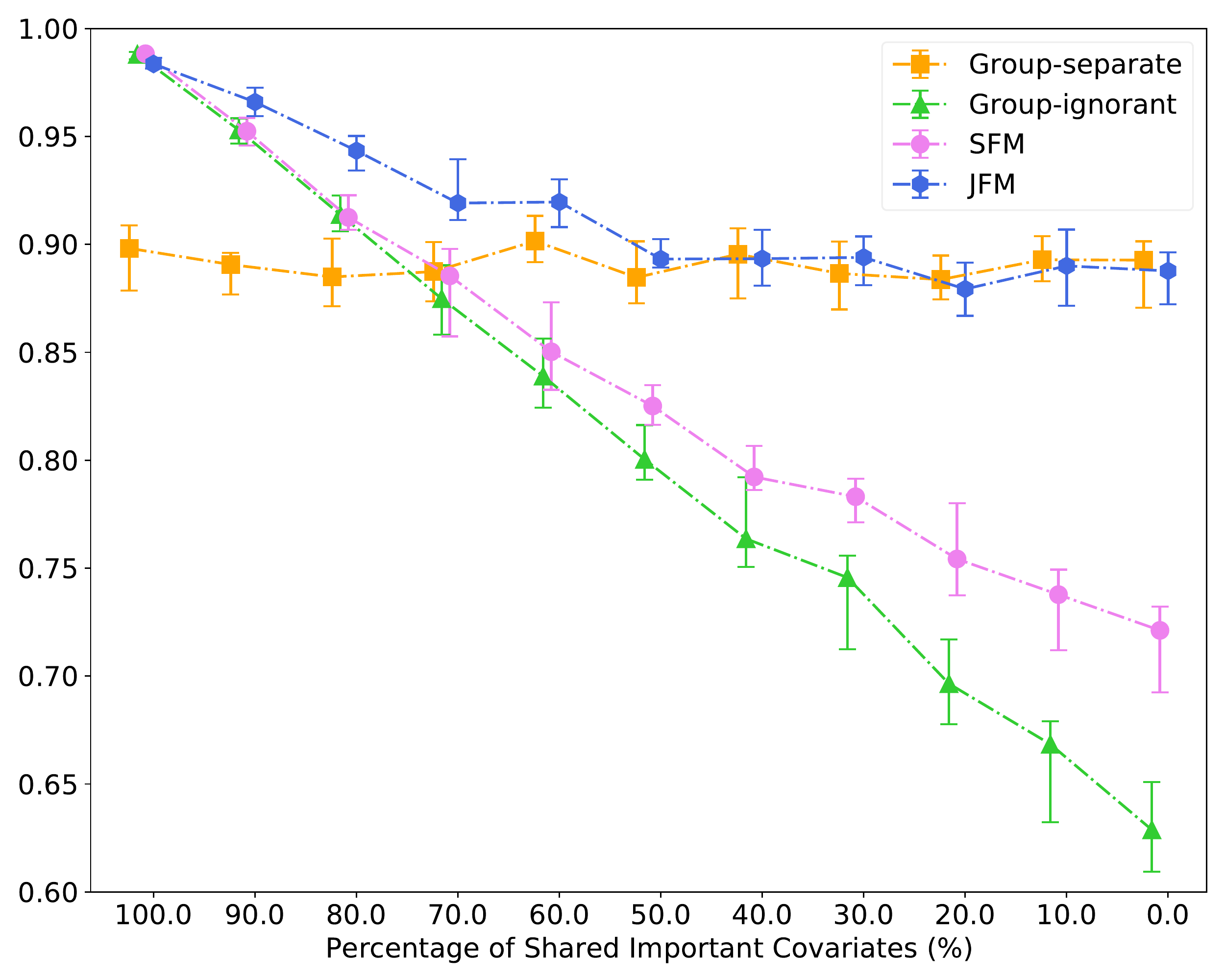}}
    \hspace{15pt} 
    \subfloat[AUC of the Over-represented Group]{\includegraphics[width=0.4\linewidth, keepaspectratio]{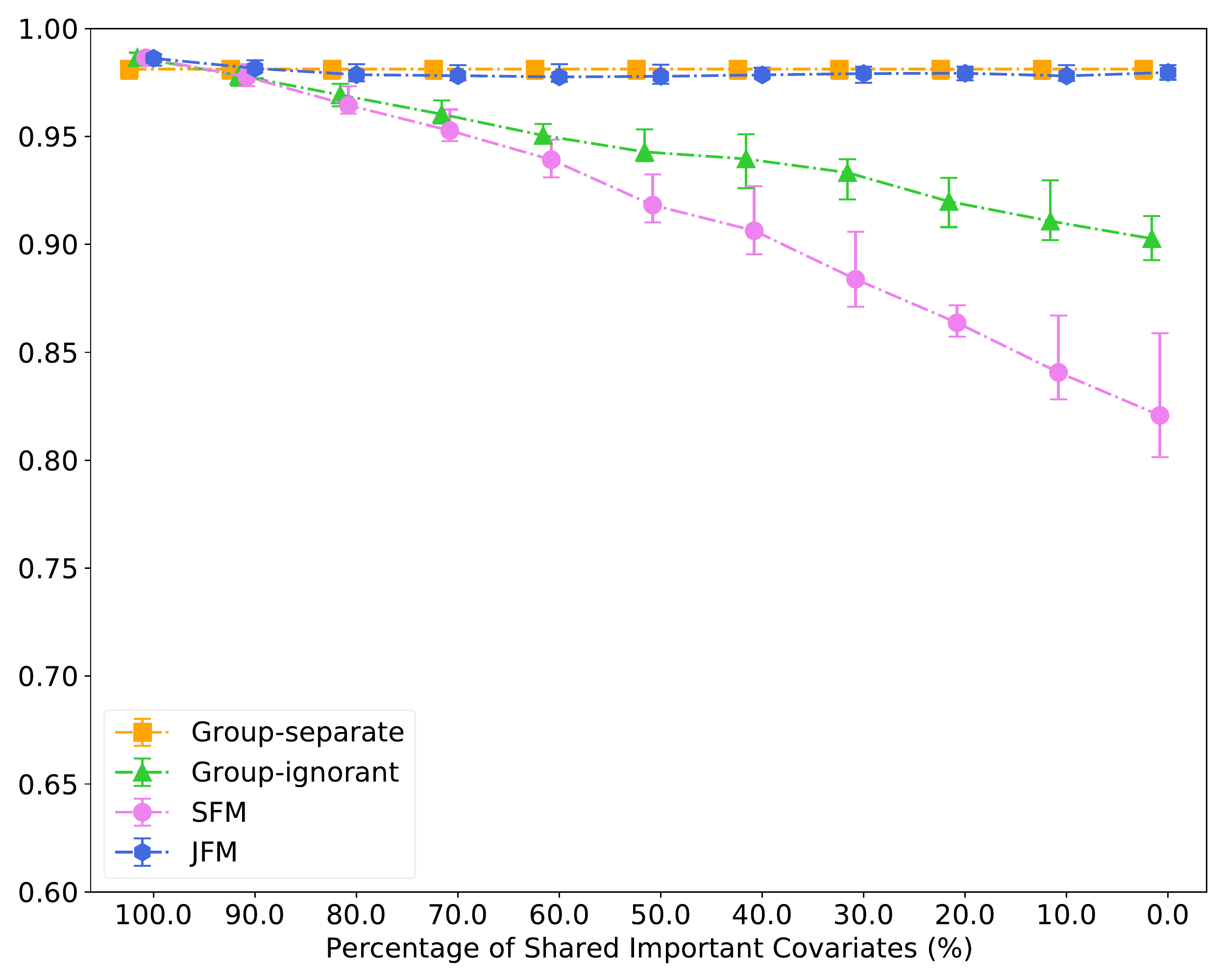}}
    \\
    \subfloat[Overall AUC]{\includegraphics[width=0.4\linewidth, keepaspectratio]{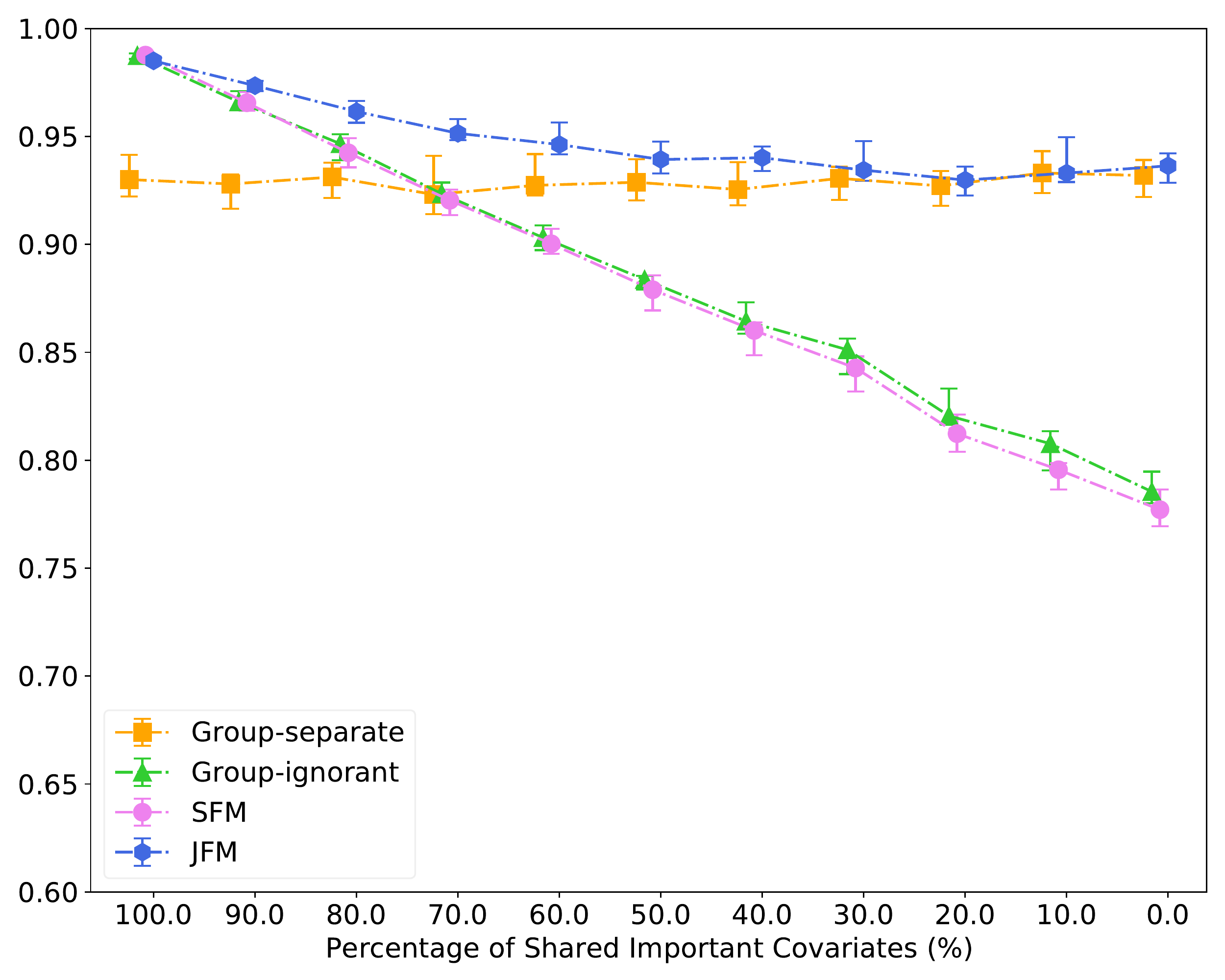}}
    \hspace{15pt} 
    \subfloat[Disparity of AUC]{\includegraphics[width=0.4\linewidth, keepaspectratio]{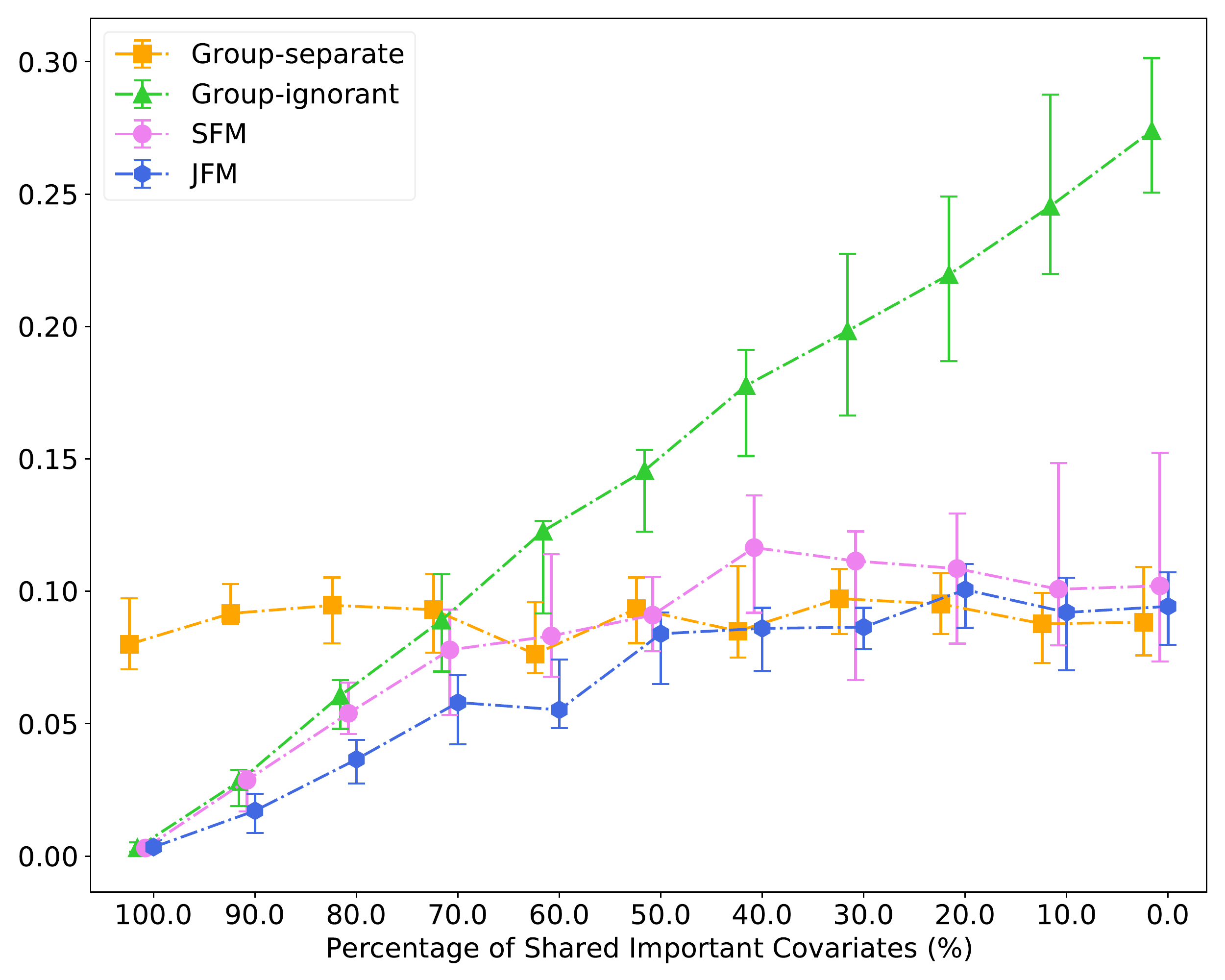}}
    \caption{Experimental Results for Scenario 1}
    \label{fig:sim1}
\end{figure}

For Scenario 1, Figure \ref{fig:sim1}(a) displays the estimated AUC for the under-represented group versus the proportion of shared features in the two groups. The AUCs of the under-represented group from the JFM, SFM, and group-ignorant models improved as the proportion of shared features increased. The SFM and group-ignorant models were highly sensitive to the percentage of shared non-zero features as they both estimate a single set of parameters for both groups. In contrast, the JFM showed consistently higher AUC than the other three methods. When the proportion of shared features is high, the JFM estimated higher AUCs and smaller variances than those from the group-separate model. The JFM's overall AUC performance was similar to that of the SFM and the group-ignorant model. When the proportion of shared features is low, the JFM estimated higher AUCs than the SFM and the group-ignorant model, and showed similar AUC to the group-separate model.
Figure \ref{fig:sim1}(b) displays the estimated AUC for the majority group against the proportion of shared features in the two groups. The JFM was robust in achieving comparable AUC to that of the group-separate model. The SFM and group-ignorant models were highly sensitive to the percentage of shared features for the majority group, with lower AUCs when the proportion of shared parameters is low.

Figure \ref{fig:sim1}(c) displays the estimated overall AUCs, and Figure \ref{fig:sim1}(d) displays the group disparity of the AUCs from the four approaches. 
In particular, although both the SFM and JFM achieved smaller disparities than the group-ignorant model in Figure \ref{fig:sim1}(d), Figure \ref{fig:sim1}(a-b) underscores the limitation of the SFM with no flexibility of group-specific parameters. When the true models of the two groups are different, the SFM often achieved better parity by compromising the performance for the majority group. These figures together demonstrate that the JFM achieves fair prediction performance robustly across the range of possible  proportions of shared features between groups, by training the classifiers jointly with a flexible parameterization. Web Figure 5(a) through 5(d) compares the average of prediction TPR and TNR and disparity in TPR and TNR differences of the four methods. The patterns are similar to those observed for AUCs.
Coefficient MSEs are presented in Figure \ref{fig:sim-coef}(a-b) for the minority and majority groups respectively. Under varying proportions of the shared parameters, JFM achieves the smallest MSEs across different methods for the minority group and similar MSEs to the group-separate model for the majority group, demonstrating its improvement for parameter estimation especially for the minority group. Web Figure 14(a) through 14(f) compares the variable selection TPRs and TNRs of the four methods. The patterns are similar to MSEs.

\begin{figure}[!t]
    \centering
    \subfloat[AUC of the Under-represented Group]{\includegraphics[width=0.4\linewidth, keepaspectratio]{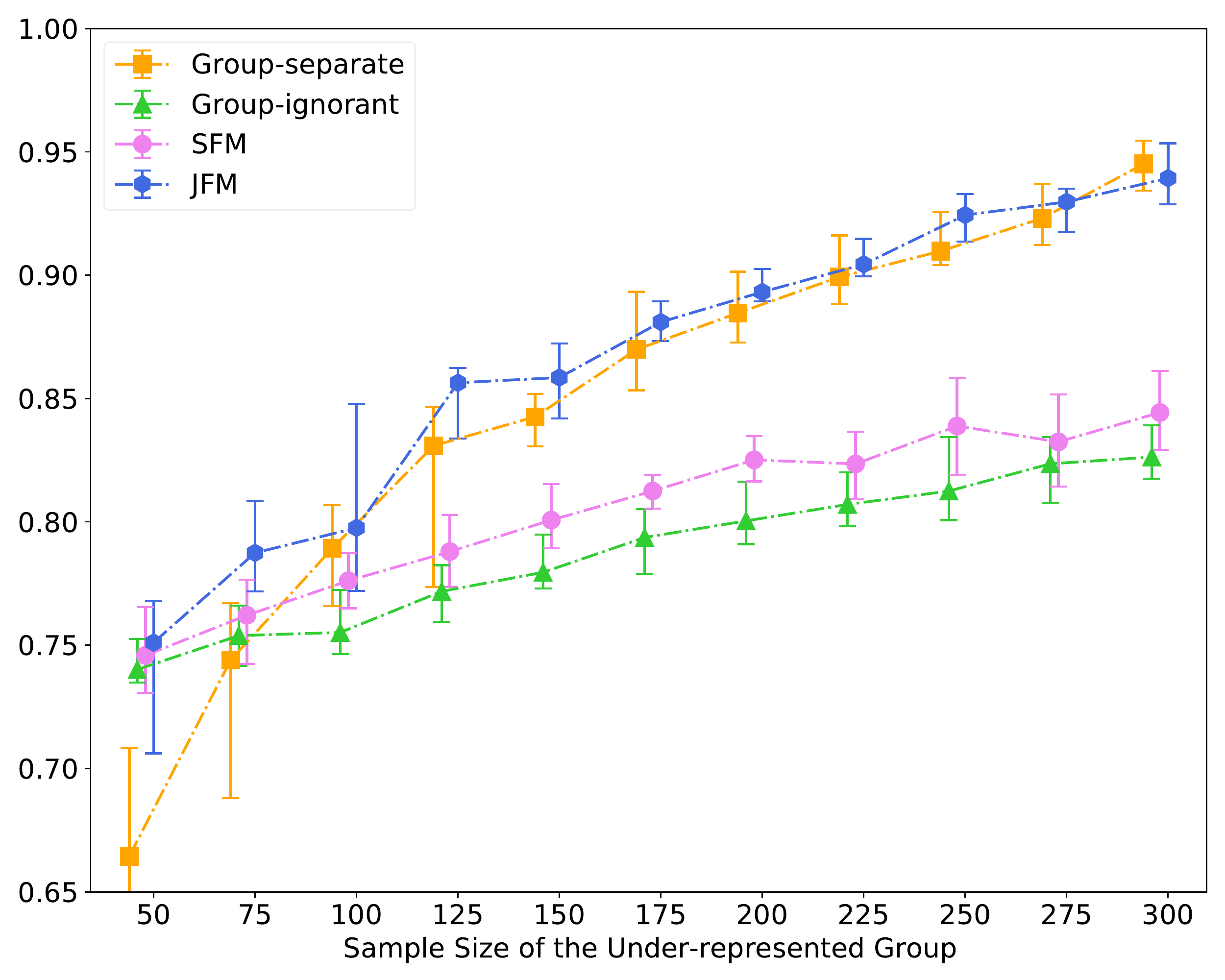}}
    \hspace{15pt} 
    \subfloat[AUC of the Over-represented Group]{\includegraphics[width=0.4\linewidth, keepaspectratio]{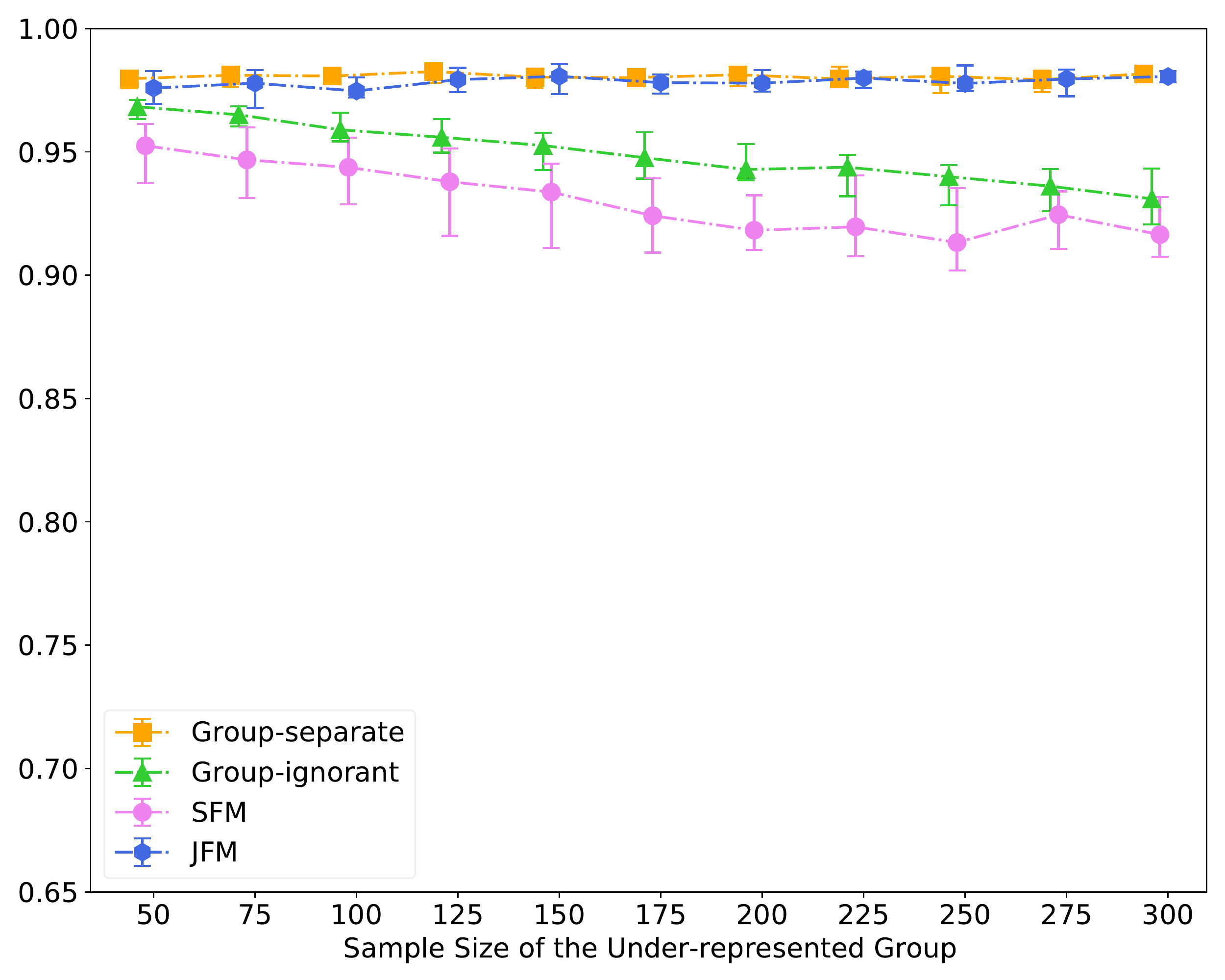}}
    \\
    \subfloat[Overall AUC]{\includegraphics[width=0.4\linewidth, keepaspectratio]{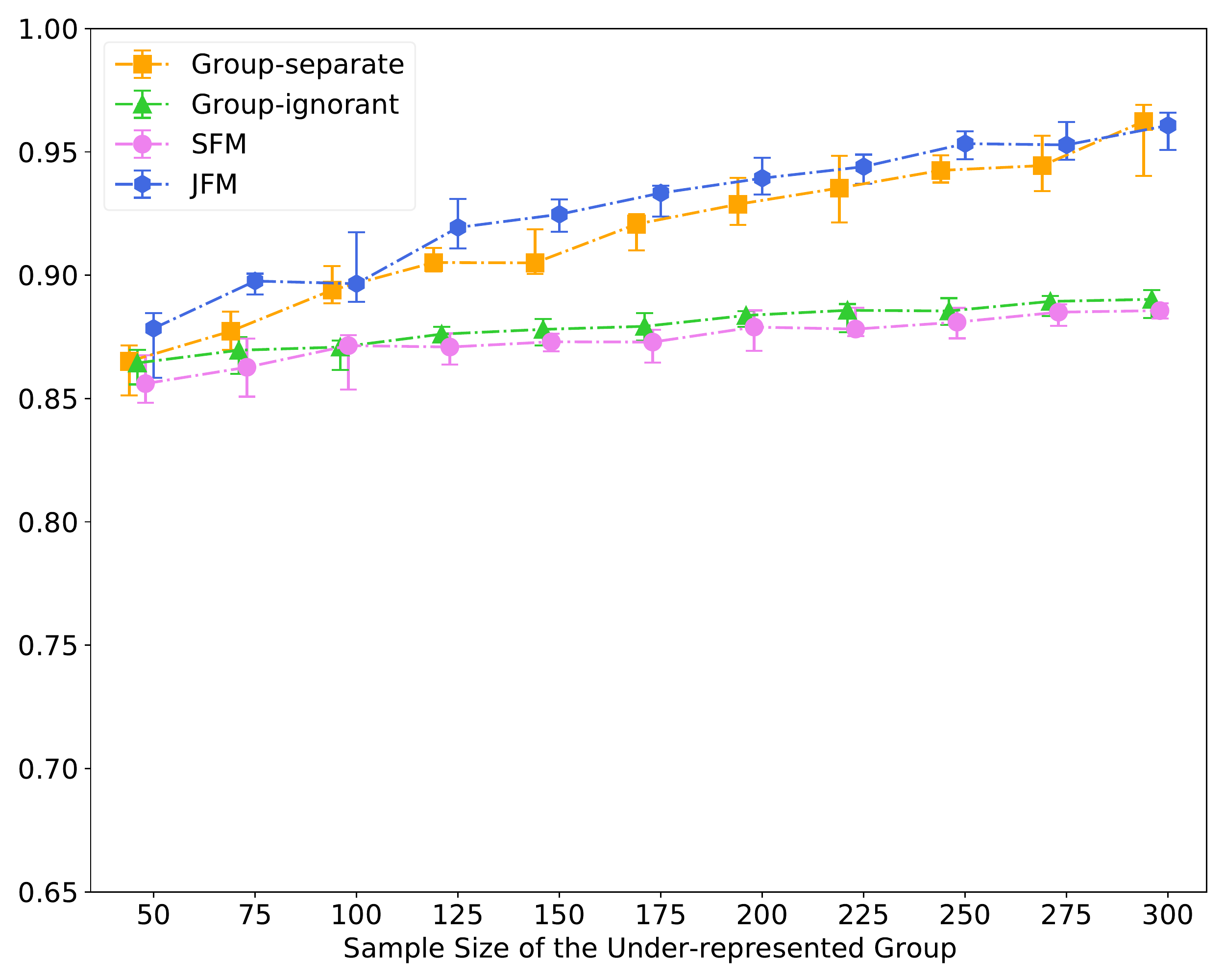}}
    \hspace{15pt} 
    \subfloat[Disparity of AUC]{\includegraphics[width=0.4\linewidth, keepaspectratio]{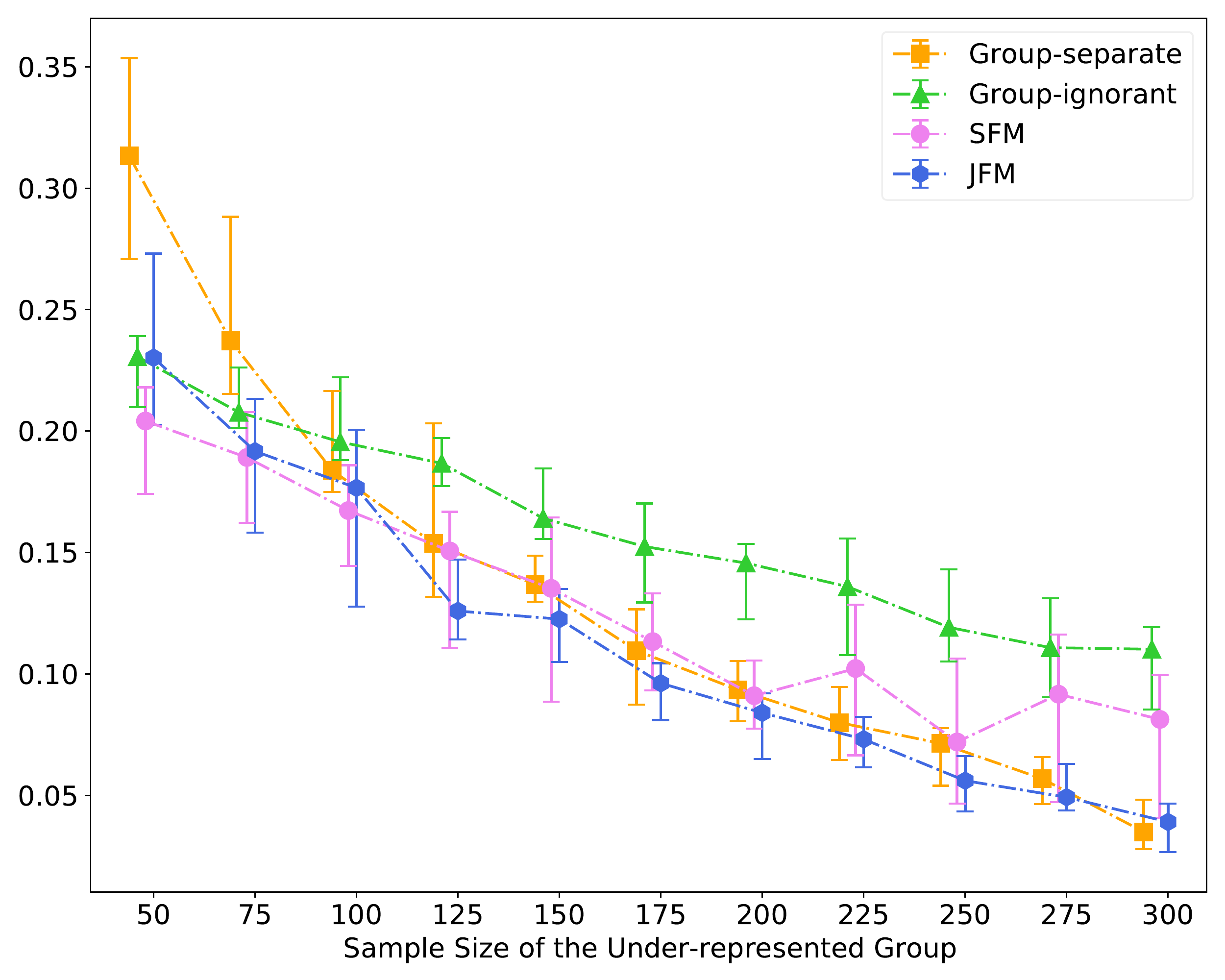}}
    \caption{Experimental Results for Scenario 2}
    \label{fig:sim2}
\end{figure}

Figure \ref{fig:sim2} displays the performance of the four methods as a function of the sample size of the under-represented group, with other settings fixed. 
In Figure \ref{fig:sim2}(a), the AUCs of the under-represented group from all models improved as  sample size of that group increased. The JFM showed consistently higher AUCs and smaller variances than those from all the other models. JFM outperforms the other models the most when the minority group's sample size is small, showing the benefits of borrowing information between groups in situations with unbalanced sample sizes.
Figure \ref{fig:sim2}(b) illustrates that the AUC of the majority group was not impacted for the JFM and group-separate methods. However, the AUC of the majority group decreased as the sample size of the under-represented group increased for the SFM and the group-ignorant models. This decrease highlights an undesirable aspect of these two methods, namely, they compromise accuracy by estimating a single set of classifier parameters. 
Figure \ref{fig:sim2}(c) and \ref{fig:sim2}(d) illustrates that the JFM achieves overall satisfactory AUCs and parity between groups across varying sample sizes of the under-represented group. Web Figure 6 compares the average of TPR and TNR and disparity of TPR and TNR of the four methods.
 In addition, the JFM substantially reduced coefficient MSEs (Figure \ref{fig:sim-coef}(c)) under all simulated sample sizes for the minority group compared to all competing methods. The MSEs were similar between the JFM and the group separate model for the majority group, and both were lower than the MSEs of the group ignorant and SFM models.

\begin{figure}[!t]
    \centering
    \subfloat[AUC of the Under-represented Group]{\includegraphics[width=0.4\linewidth, keepaspectratio]{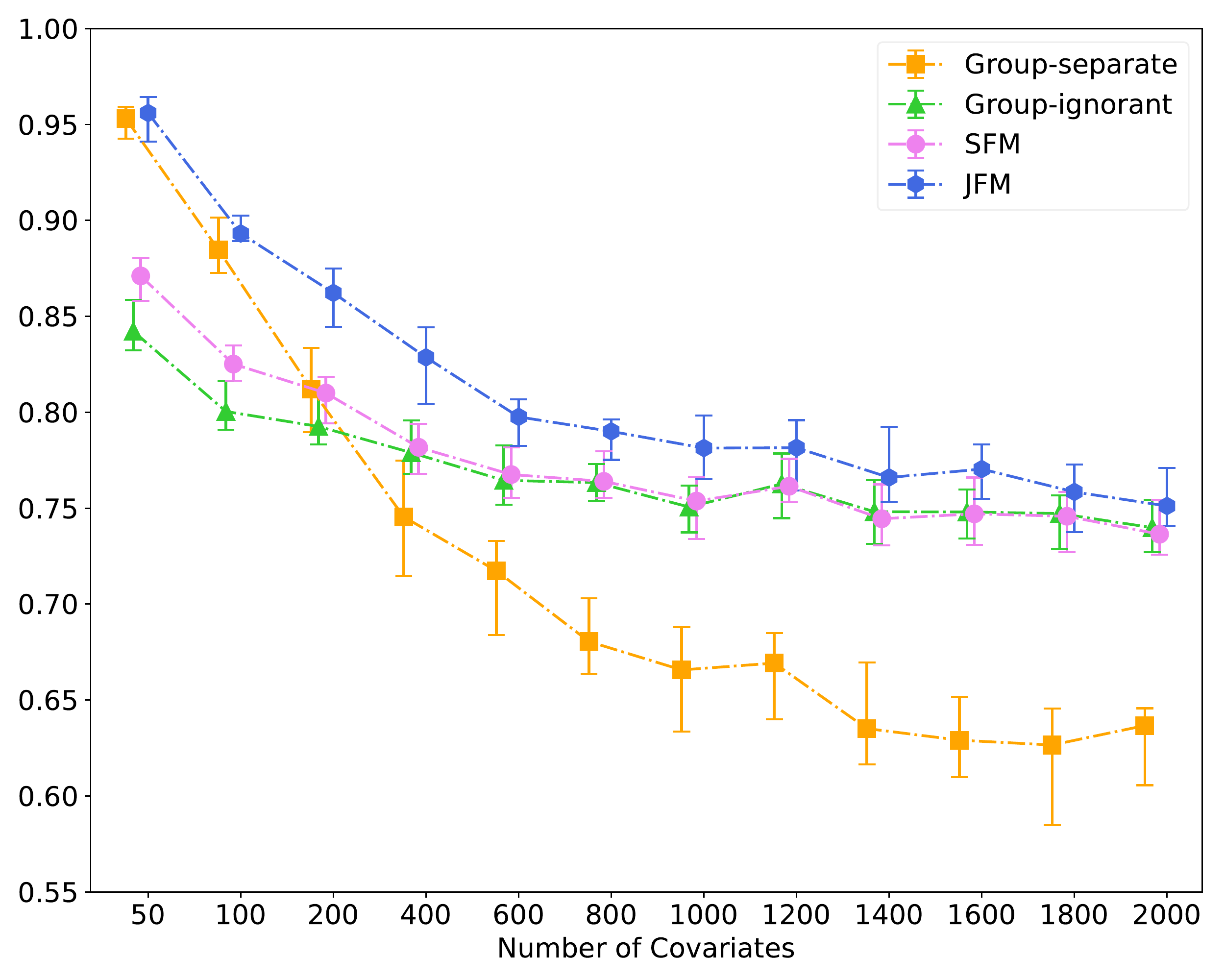}}
    \hspace{15pt} 
    \subfloat[AUC of the Over-represented Group]{\includegraphics[width=0.4\linewidth, keepaspectratio]{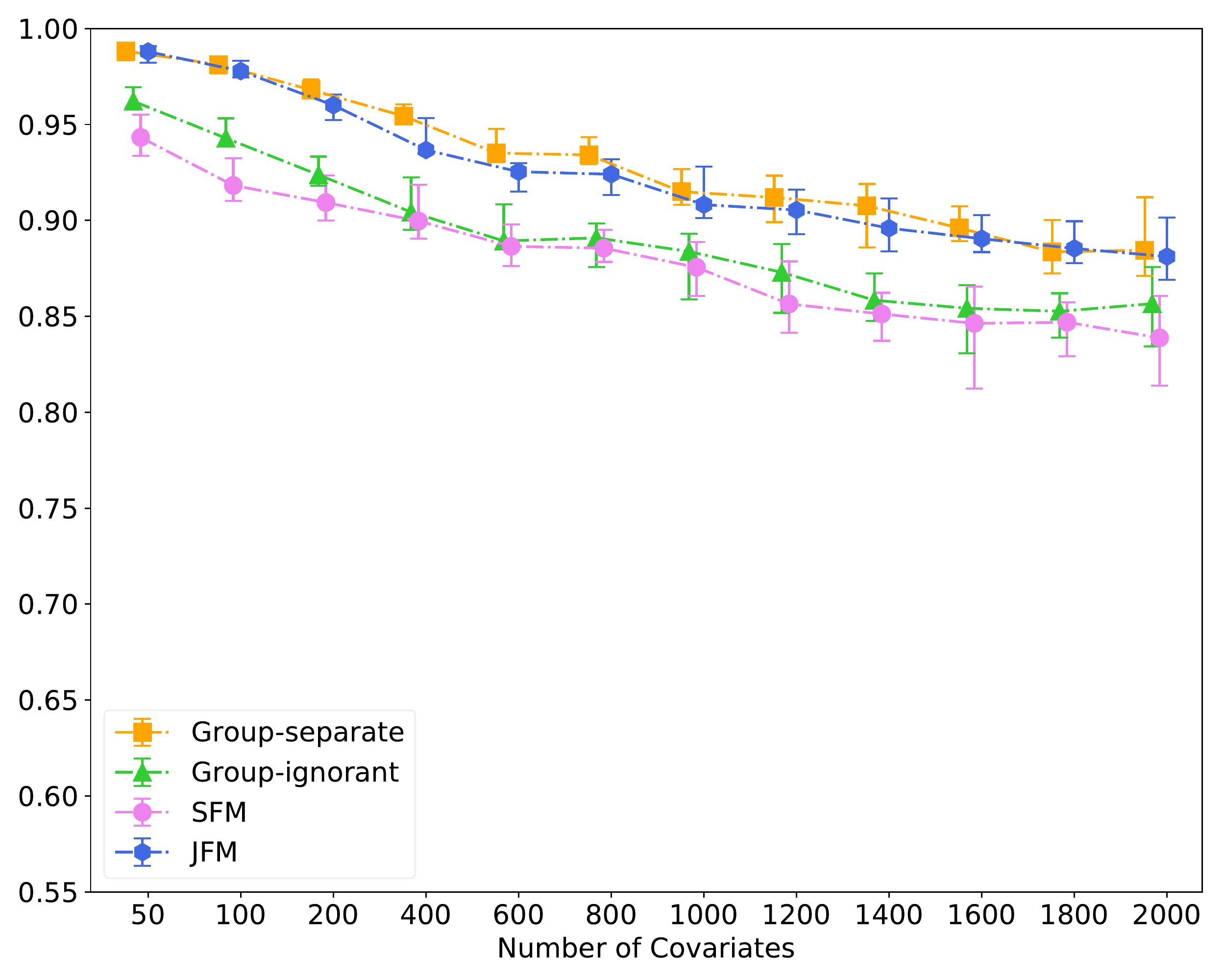}}
    \\
    \subfloat[Overall AUC]{\includegraphics[width=0.4\linewidth, keepaspectratio]{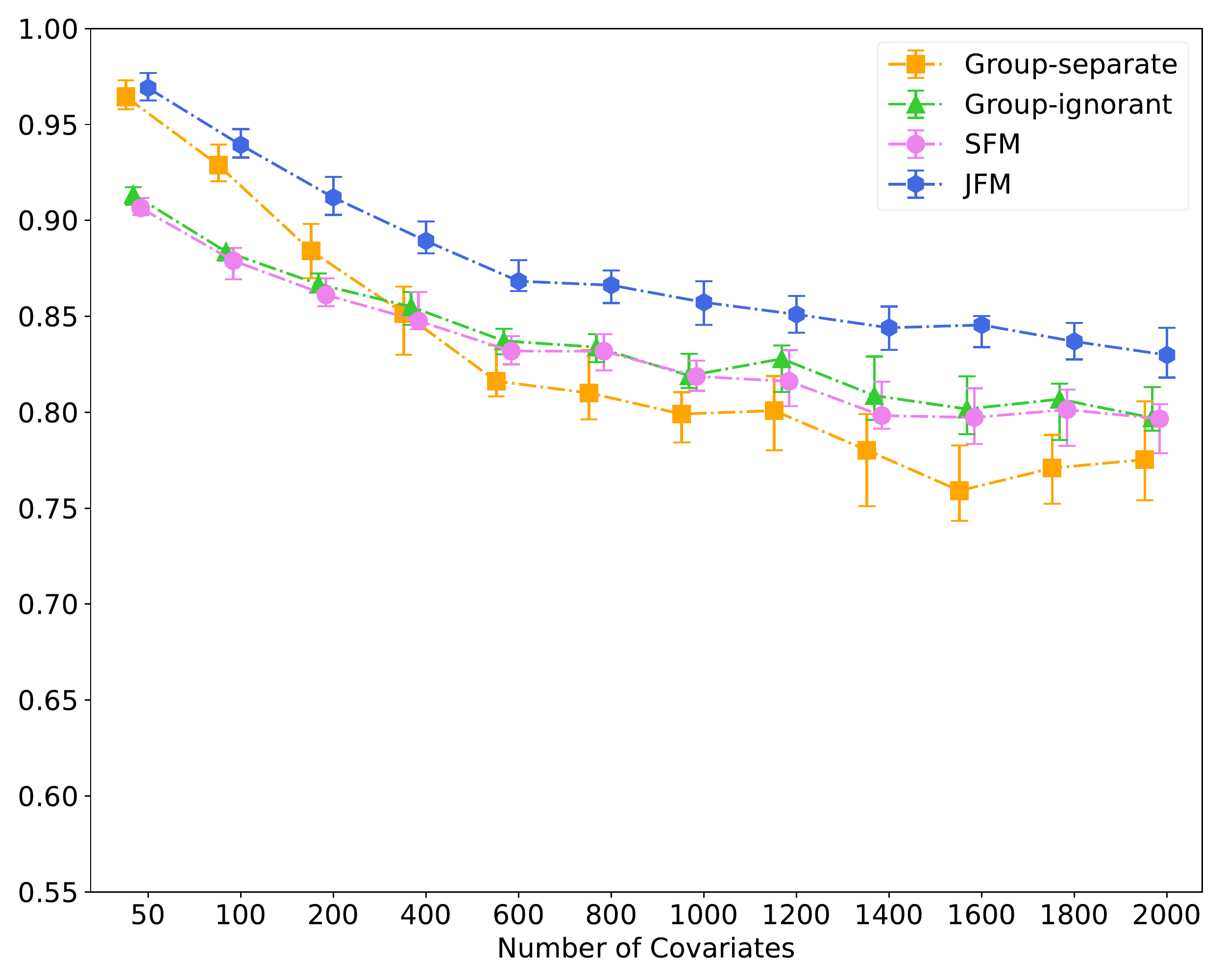}}
    \hspace{15pt} 
    \subfloat[Disparity of AUC]{\includegraphics[width=0.4\linewidth, keepaspectratio]{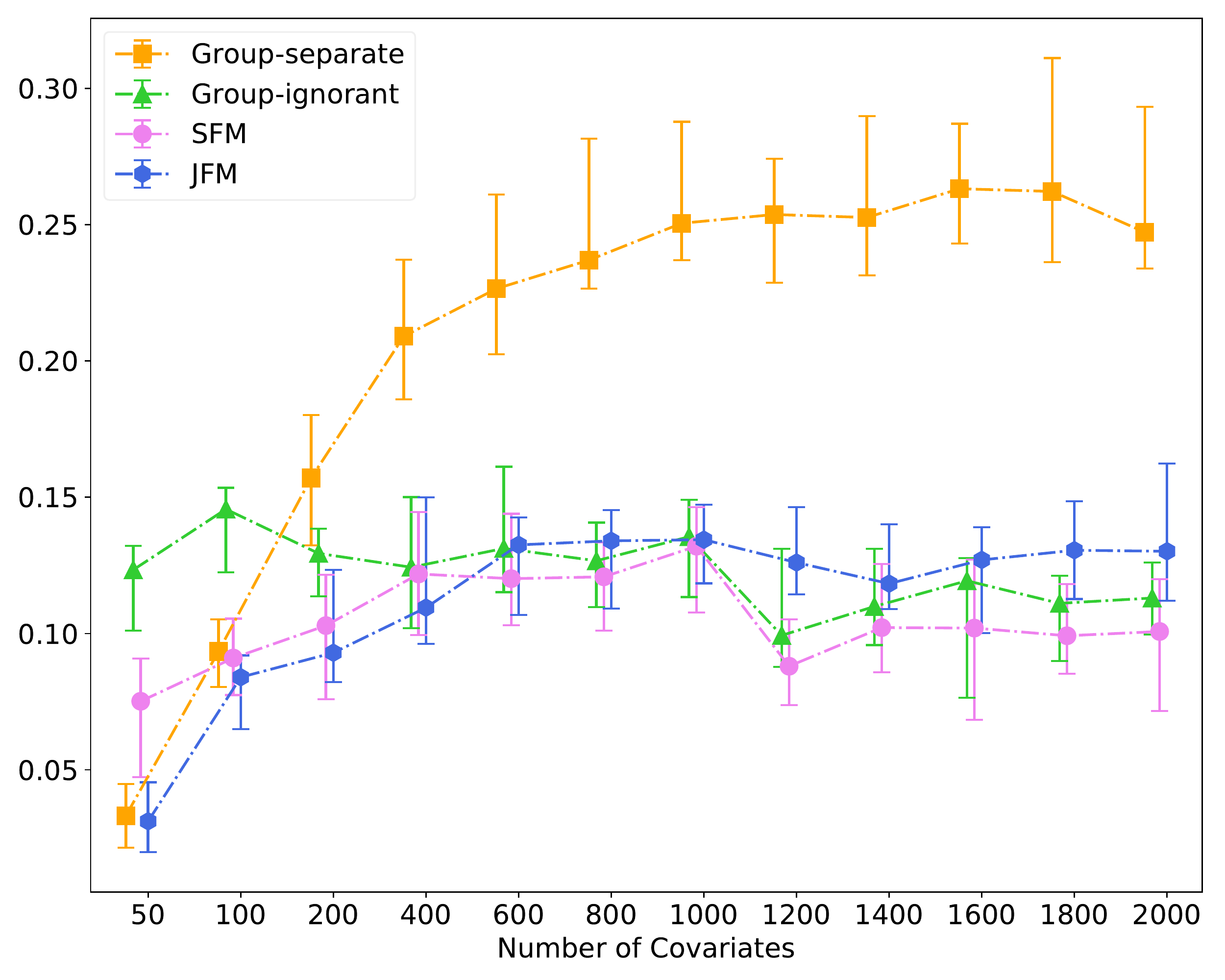}}
    \caption{Experimental Results for Scenario 3}
    \label{fig:sim3}
\end{figure}

\begin{figure}[!t]
    \centering
    \subfloat[Scenario 1 Coefficients MSEs of the Under-represented Group]{\includegraphics[width=0.45\linewidth, keepaspectratio]{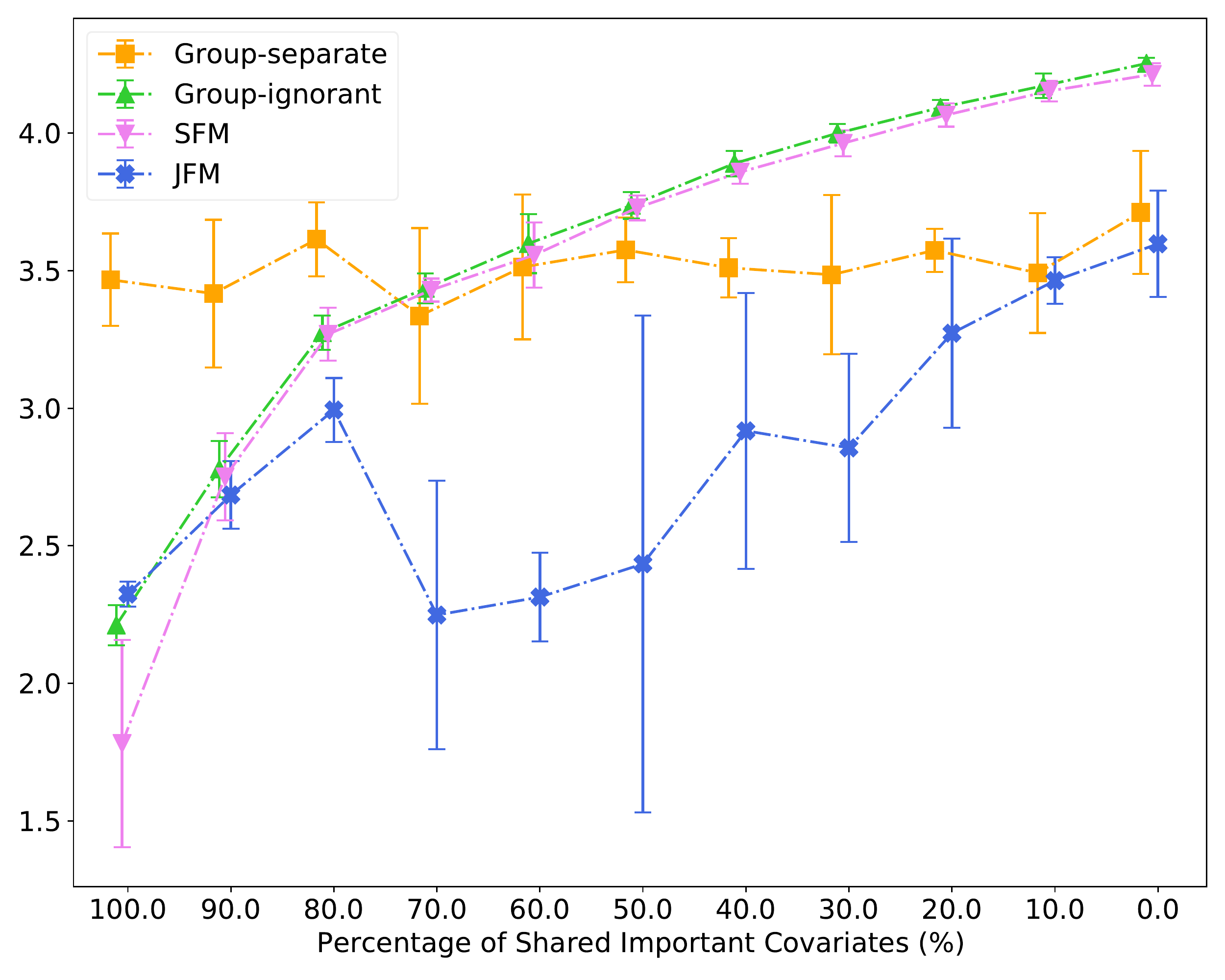}}
    \hspace{5pt}
    \subfloat[Scenario 1 Coefficients MSEs of the Over-represented Group]{\includegraphics[width=0.45\linewidth, keepaspectratio]{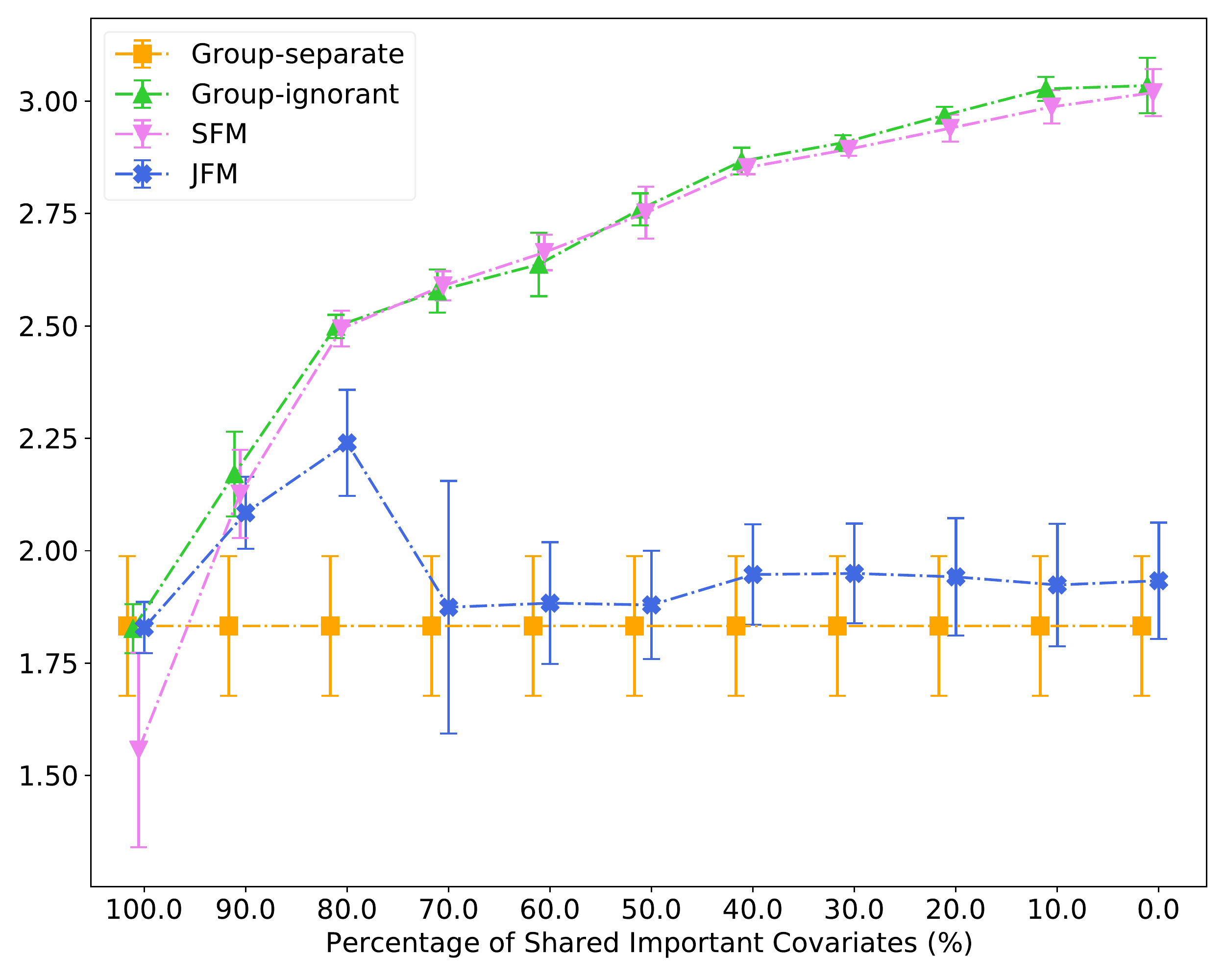}}
    \\
    \subfloat[Scenario 2 Coefficients MSEs of the Under-represented Group]{\includegraphics[width=0.45\linewidth, keepaspectratio]{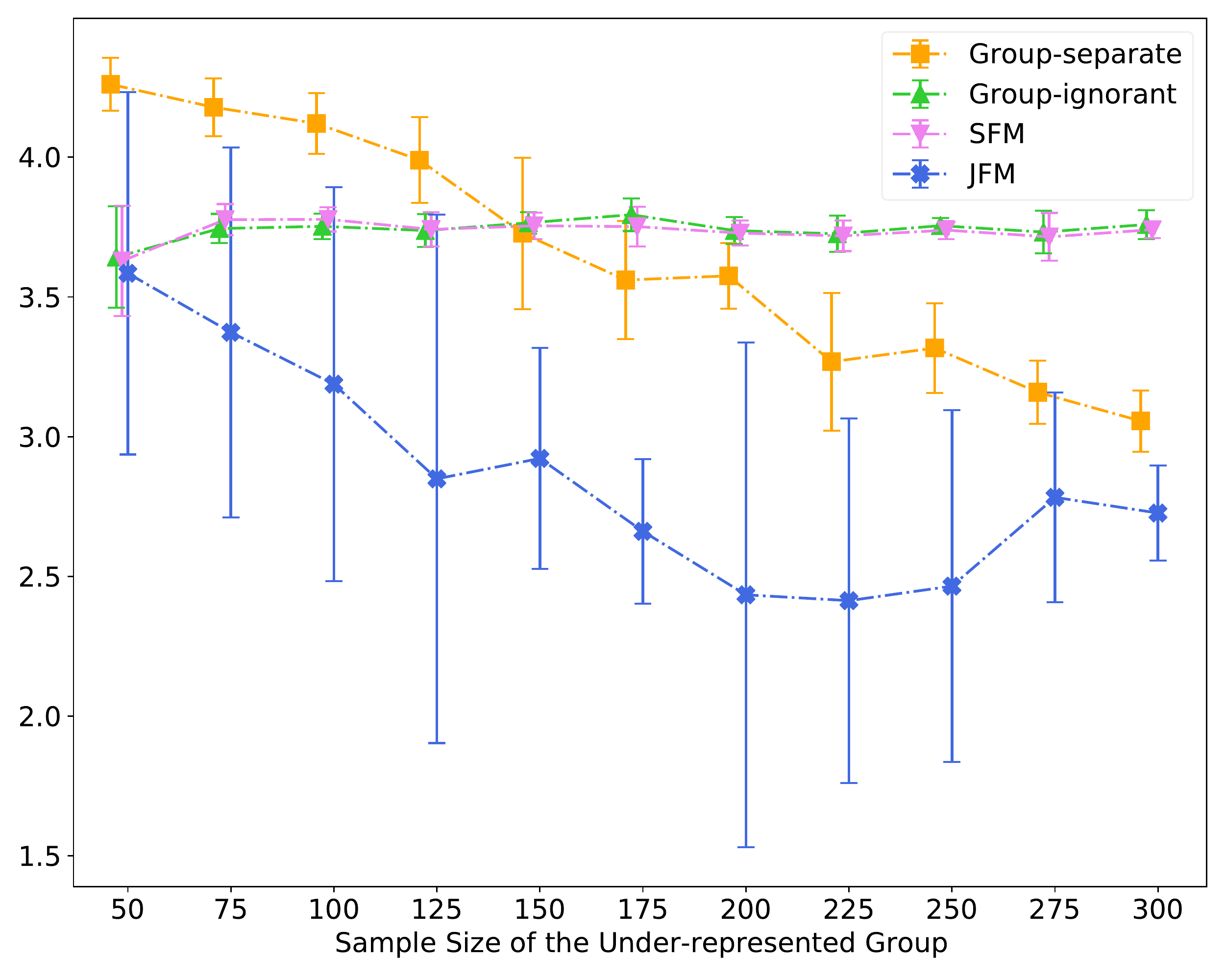}}
    \hspace{5pt}
    \subfloat[Scenario 2 Coefficients MSEs of the Over-represented Group]{\includegraphics[width=0.45\linewidth, keepaspectratio]{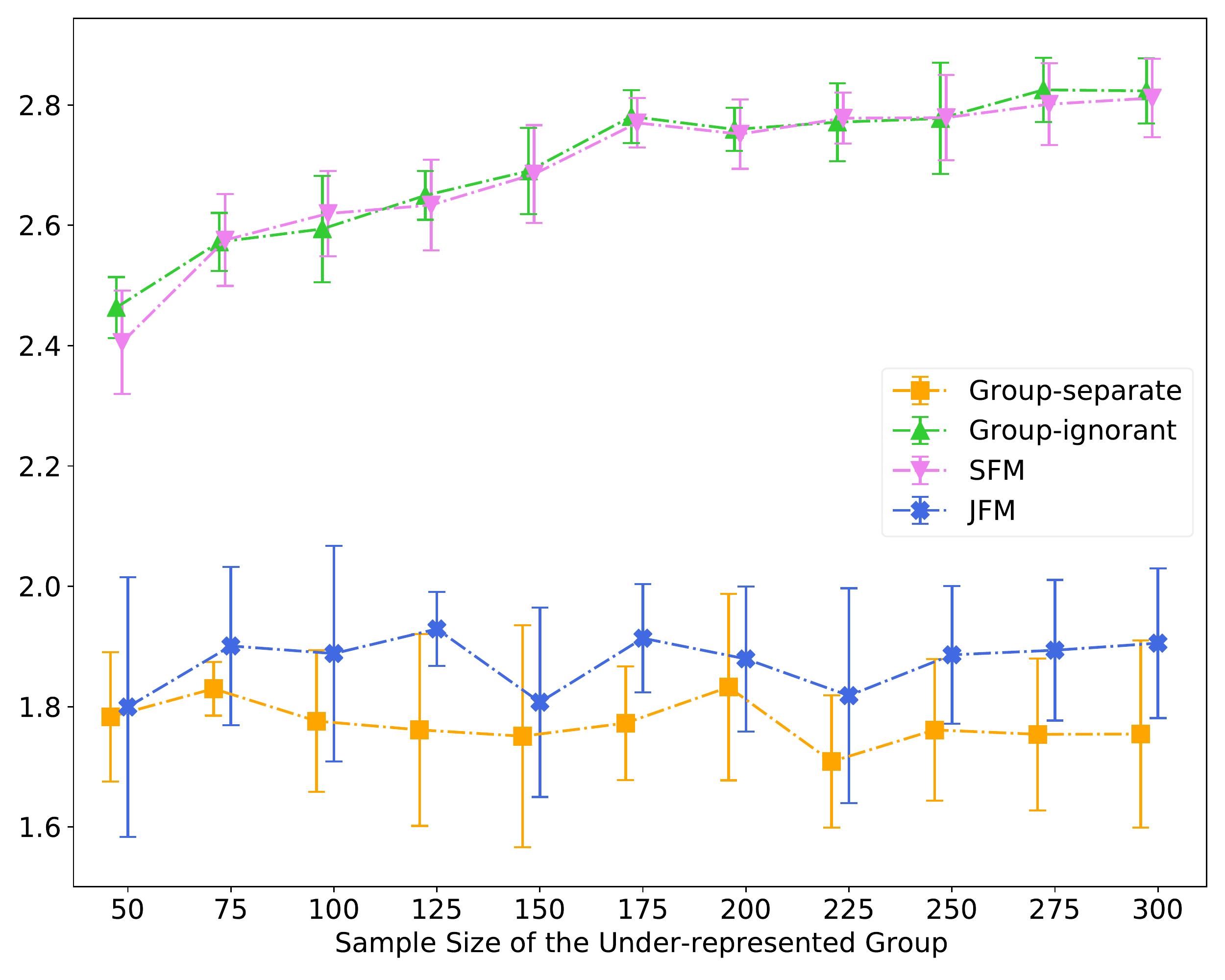}}
    \\
    \subfloat[Scenario 3 Coefficients MSEs of the Under-represented Group]{\includegraphics[width=0.45\linewidth, keepaspectratio]{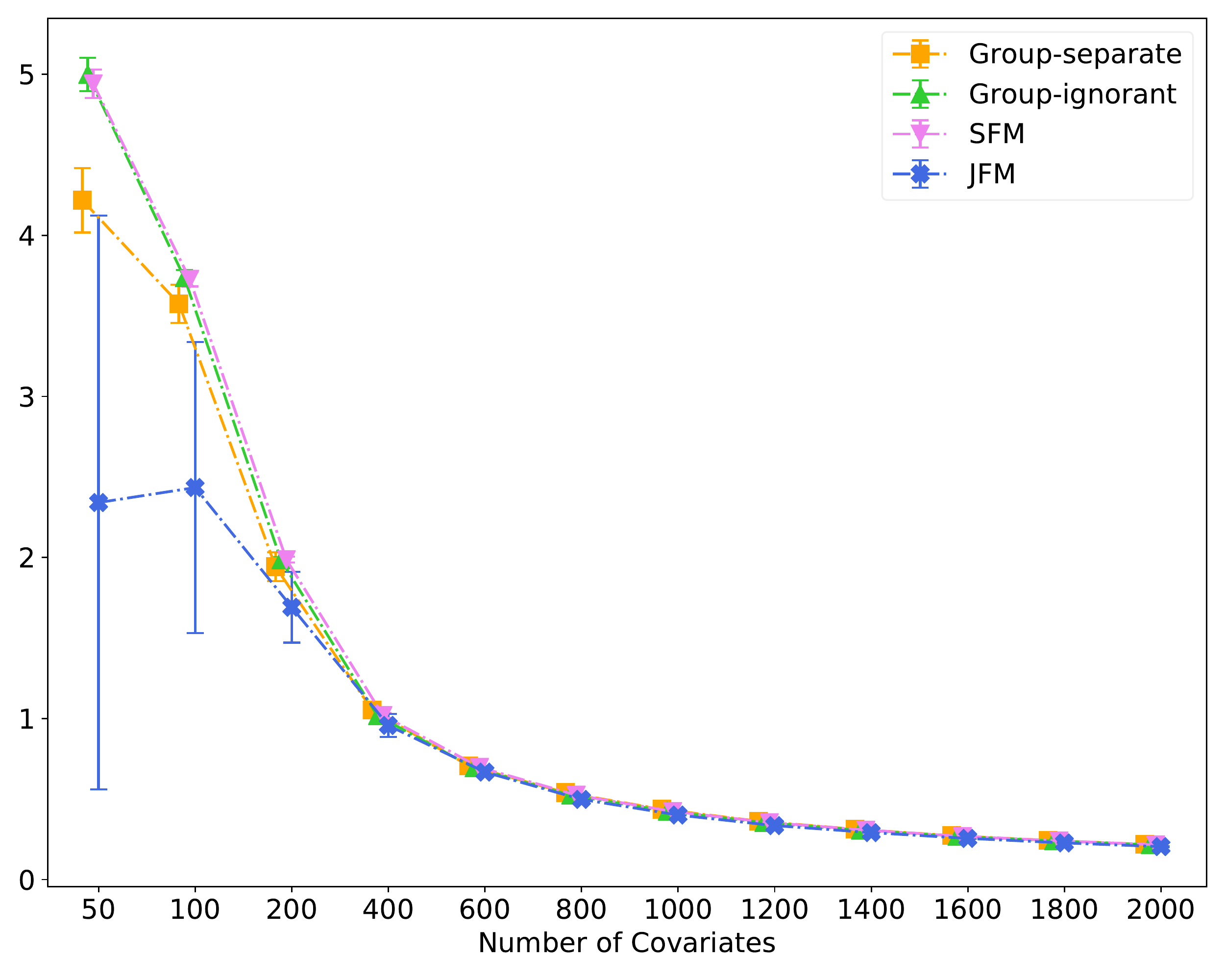}}
    \hspace{5pt}
    \subfloat[Scenario 3 Coefficients MSEs of the Over-represented Group]{\includegraphics[width=0.45\linewidth, keepaspectratio]{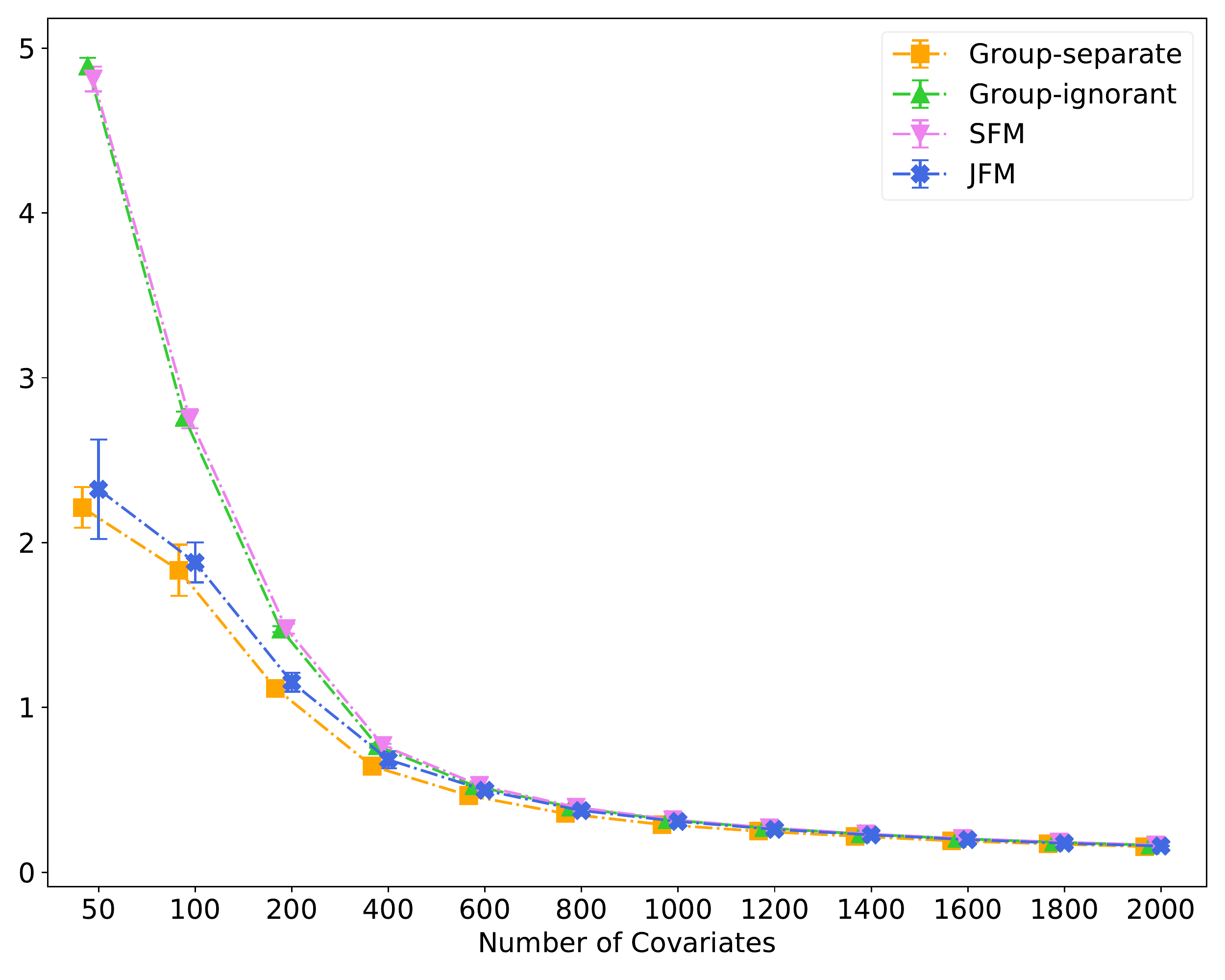}}
    \caption{Coefficients MSEs for Scenario 1 -- 3}
    \label{fig:sim-coef}
\end{figure}

Figure \ref{fig:sim3} displays the performance of the four methods while varying the number of features from 200 to 2000, and holding the number of associated features constant at 40. It demonstrates that the JFM method in going from low dimensional to high dimensional settings can maintain  overall satisfactory prediction performances and parity between groups. Web Figure 11 displays the performance of the four methods while varying the number of features from 200 to 2000, and setting the number of associated features to a fixed proportion of the total number of features. The resultant patterns are similar to Figure \ref{fig:sim3}.
The JFM  consistently had the smallest MSEs under all simulated numbers of covariates for both the minority and majority groups. In particular, the JFM showed the largest reduction of MSEs with small number of covariates.

We investigated the empirical computational complexity of the JFM approach  as a function of increasing numbers of features and sample sizes in Web Appendix 3. Web Figure 1 shows that the JFM computation time is approximately $\mathcal{O}(p^{1.5})$ and $\mathcal{O}(n)$. This matches our theoretical analysis on complexity since the per-iteration complexity is $\mathcal{O}(np)$ and the rate of convergence is $\mathcal{O}(p^{0.5})$. Details are presented in Web Appendix 6.
We also implemented an alternative JFM algorithm using a group lasso similarity penalty (referred to as JFM-Group, Web Figure 15), and compared its performance with JFM-Fusion (presented above). The results showed largely similar performances. The JFM-Fusion showed slightly better predictive performance than the JFM-Group when the two groups share more than 80\% of the features, which is in line with previous reports that the fusion penalty term more aggressively enforces similarities across groups \citep{yuan2006model, obozinski2010joint, danaher2014joint}.

\section{COVID-19 Risk Prediction Case Study}
\label{sec:app}
\begin{figure}[!t]
    \centering
    \includegraphics[width=0.9\linewidth, keepaspectratio]{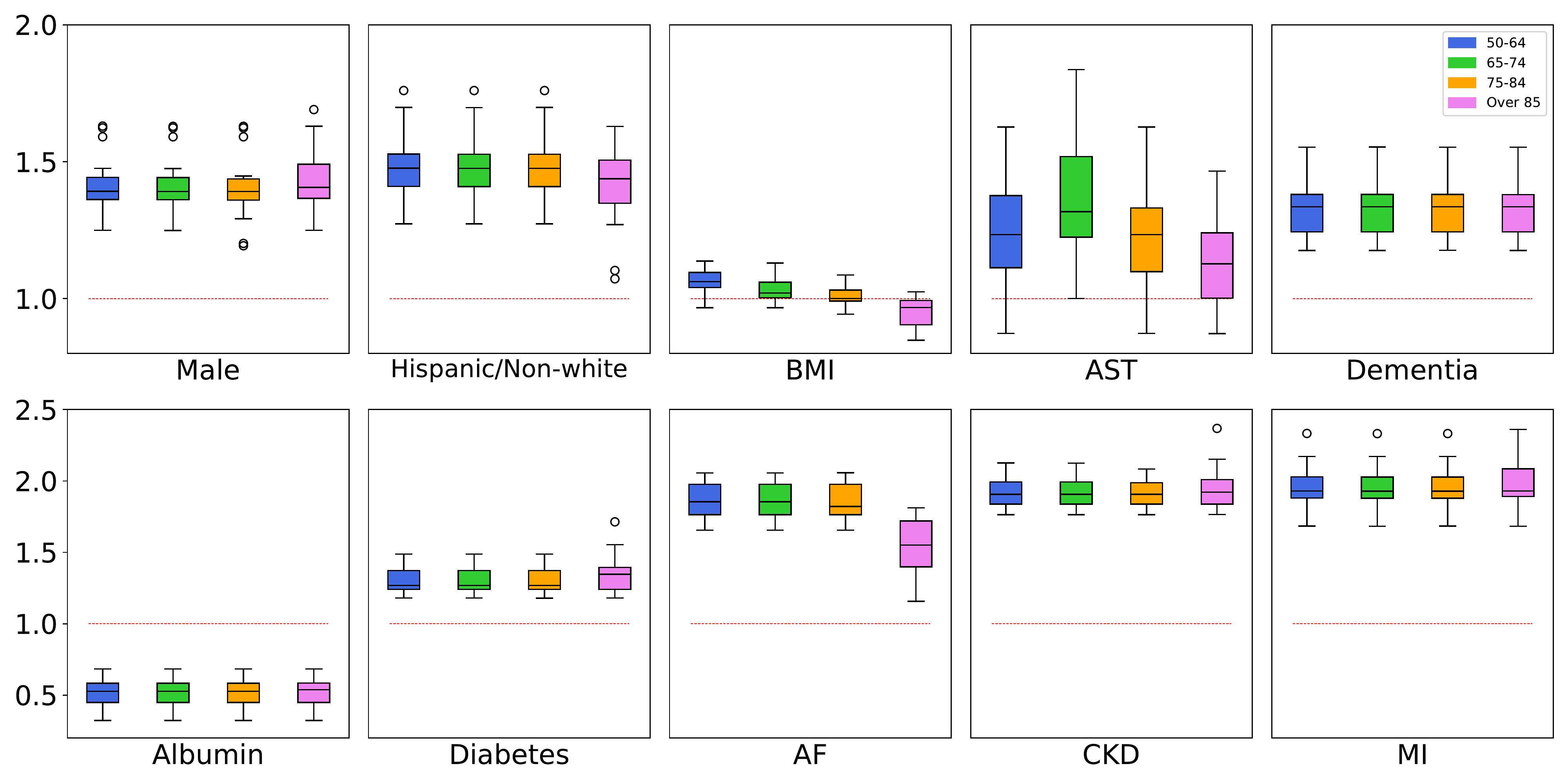}
    \caption{Estimated Odds Ratios for COVID-19 Dataset}
    \label{fig:case-study-odds-ratios}
\end{figure}

We applied the JFM, in comparison with other methods, to predict mortality related to COVID-19 from patients' routine ambulatory encounters and laboratory records prior to COVID-19 infection, with the goal of better stratification of patient risk for clinical management.
We used a retrospective EHR dataset of 11,594 patients of age 50+ with laboratory-confirmed COVID-19 at New York University Langone Health (NYULH) from March 2020 to February 2021. 
Among the 11,594 patients, 1,242 (10.7\%) died of COVID-19. The patients were divided into four groups by their age at the time of COVID-19 diagnosis: 50-64, 65-74, 75-84, and 85+ with $5,905$ (50.9\%), $2,946$ (25.4\%), $1,814$ (15.6\%), and $929$ (8.0\%) patients, respectively. The observed mortality rates were 4.44\%, 11.17\%, 18.96\% and 33.05\%, respectively. Candidate features ($p = 82$) included demographic variables, such as age, sex, race/ethnicity, smoking status, body mass index (BMI); common chronic disease history such as diabetes, dementia, chronic kidney diseases (CKD); Myocardial Infarction (MI) \& Atrial Fibrillation (AF); and routinely collected  laboratory markers, such as lipid panels, blood panels, albumin, creatinine, aspartate aminotransferase (AST) etc. obtained from patients routine ambulatory histories before their COVID-19 infections. 
To build the prediction models, we randomly split the dataset into training ($n=8,115$, 70\%) and testing ($n=3,479$, 30\%) sets. We first standardized all features to zero-mean and unit variance. Five-fold cross-validation was conducted on the training set to determine the hyperparameters for each model. Hyperparameters for the group-separate and group-ignorant models were selected to maximize the groupwise AUCs and the overall AUC, respectively, while those for the SFM and JFM were determined to maximize the harmonic mean of groupwise AUCs. Subsequently, we trained the final models with the optimal hyperparameters using the entire training set and applied the final models to the testing dataset to demonstrate their predictive performance. We repeated the training/testing split 10 times and averaged the performances across the 10 splits. 
Table \ref{table:case-study} presents the AUCs and the averages of TPR and TNR of the four methods for each age group. The JFM performed better across all age groups than the separate model did, demonstrating that joint modeling yields higher efficiency. Compared with the group ignorant model, the JFM performed better in the three older age groups, with comparable AUC for the 50-64 age group, which resulted in smaller disparities in prediction performance overall. This phenomenon supports the observed pattern in simulation studies that the JFM reduced disparities in prediction performances without impacting those from the majority groups. In contrast, the SFM tended to reduce prediction disparities by lowering the performances for the majority groups. 
Figure \ref{fig:case-study-odds-ratios} presents the boxplots of odds ratios (ORs) of selected demographic and clinical features estimated by the JFM. 
These results support the hypothesis that some features have common associations between groups, and some have group-specific ORs. 
For example, the decreasing OR estimates of BMI along age-groups confirmed the prior hypothesis that the association between BMI and COVID-19 mortality is heterogeneous between age-groups. In the JFM estimates, BMI is positively associated with higher risks of COVID-19 mortality for patients younger than 75, but with smaller and even reversed ORs in the oldest age groups.   
For older adults, higher BMIs are often associated with greater energy stores and a better nutritional state overall, which is beneficial for patients' survival outcomes when infected by COVID-19. 
The proportion of underweight patients (BMI$<$18) increased from 0.6\% in the age group 50-64 to 5.5\% in the age group 85+. 
The underweight status, often a proxy of frailness, has been repeatedly reported as a strong risk factor of COVID-19-induced multi-organ failure and mortality in older patients \citep{TEHRANI2021415}.
On the other hand, the JFM can improve efficiencies for covariates with rare prevalence in a subgroup. For instance, dementia has been reported as a risk factor with COVID-19 mortality. In the group-separate model, dementia was insignificant in patients aged 50-64, mainly due to its low prevalence in this group (0.6\%). In contrast, dementia was significantly associated with mortality in all age groups with similar ORs in the JFM estimates.
\begin{table}[t]
    \centering
    \begin{tabular}{c|cccc|cccc}
    \hline\hline 
     \multirow{2}{*}{Models} & \multicolumn{4}{c|}{AUCs} & \multicolumn{4}{c}{Average of TPR and TNR}\\
                                 & 50-64 & 65-74 & 75-84 & Over 85 & 50-64 & 65-74 & 75-84 & Over 85 \\ \hline
         Group-separate          & 0.838    & 0.773    & 0.709    & 0.649   & 0.780    & 0.722    & 0.669    & 0.632\\
         Group-ignorant          & \textbf{0.855}    & 0.786    & 0.735    & 0.659    & \textbf{0.803}    & \textbf{0.731}    & 0.687    & 0.639\\
         SFM                     & 0.847    & 0.774    & 0.728    & 0.660   & 0.791    & 0.724    & 0.688    & 0.640\\
         JFM                     & 0.852    & \textbf{0.791}    & \textbf{0.736}    & \textbf{0.672}    & 0.794    & \textbf{0.731}    & \textbf{0.690}    & \textbf{0.659} \\ \hline\hline \\
    \end{tabular}
    \caption{Predictive performance on COVID-19 case study. The boldface items in the table represent the best performance in each column.}
    \label{table:case-study}
\end{table}

\section{Conclusions and Discussion}
\label{sec:con}
In this study we introduced a joint fairness model for jointly estimating sparse parameters, on the
basis of observations drawn from distinct but related groups, with the goal of achieving fair performances across groups. 
We employ an efficient accelerated smoothing proximal gradient algorithm to solve the joint fair objective function, which has convex penalty functions. Our algorithm is computationally tractable for high-dimensional datasets. 
Further, we presented the asymptotic distributions of $\hat{\boldsymbol{\beta}}_k$. Our JFM predictions outperform competing approaches over a range of simulations and in an example application dataset.

We note that the JFM relies on separate hyperparameters ($K+2$ hyperparameters) to control sparsity, fairness and similarity. This reliance can be viewed as a strength rather than a drawback because one can vary separately the amount of similarity, sparsity and fairness to enforce in the group-specific estimates. In situations with many groups, further assumptions can be made to reduce the number of sparsity hyperparameters (i.e. $\lambda_{\text{Sp}_{k}} = c_{k}\lambda_{\text{Sp}}$). Possible choices of $c_{k}$ include $\frac{1}{\sqrt{n_{k}}}$ so that sparsity is inversely proportional to the number of samples.

While nearly all existing fairness-aware prediction approaches estimate a single set of classifier parameters, one exception is a recent study that proposes using multi-task learning (MTL) to improve algorithm fairness \citep{10.1145/3306618.3314255}. However, MTL research focuses on joint architecture, optimization, and task relationship learning, which is a different emphasis from the proposed JFM approach to improve risk prediction performance for under-represented populations. 

Theorem 4.1 established consistency of the JFM estimators. Its asymptotic distribution needs to be further investigated to lay the foundation of its inference. Moving forward, the proposed JFM framework can be extended for time-to-event outcomes by using similar constraints to those proposed here. It can also in principle be extended to non-linear models by adding a suitable fairness penalty term to the objective function. 

Given the increasing ability to subclassify diseases according to their molecular features and the recognition that substantial heterogeneity exists in many molecular subtypes, most diseases will be eventually classified into a collection of multiple subtypes with unbalanced sample sizes. Therefore, the proposed JFM has wide application potential to improve prediction efficiencies and reduce subgroup prediction disparities beyond applications addressing gender, race/ethnicity and age disparities.

A Python package implementing the JFM is available at \\
\texttt{https://github.com/hyungrok-do/joint-fairness-model}.

\section*{Data Availability Statement}
The data that support the findings in this paper are available on request from the corresponding author. The data are not publicly available due to privacy or ethical restrictions.

\bibliography{jfm.bib}


\label{lastpage}

\newpage
\begin{algorithm}[!p]
	\caption{Accelerated Smoothing Proximal Gradient (ASPG) Algorithm for the JFM}
	\label{pseudocode:JointFair}
	\begin{algorithmic}[1]
		\State \textbf{Input:} Data $\mathbf{X}^k, \mathbf{y}^k$ for $k = 1 \dots K$, hyperparameters $\lambda_{\text{F}},\lambda_{\text{Sim}},\lambda_{\text{Sp}}$, $\epsilon$, $\mu$
        \State \textbf{Output:} $\hat{\boldsymbol{\beta}} = (\hat{\boldsymbol{\beta}^{1}}, \cdots, \hat{\boldsymbol{\beta}}^{K})$ solving the joint fairness objective function \eqref{eqn:formulation2}.
        \State \textbf{Initialize:} $\boldsymbol{\beta}^{(0)} = \mathbf{0}, ~ \boldsymbol{\gamma}^{(0)} = \mathbf{0},~ s^{(0)} = 1$
        \item $L = \frac{1}{4}\max\left\{\lambda_{\max}(\mathbf{X}^{kT}\mathbf{X}^{k}): k=1,\cdots,K\right\} + \mu^{-1}\|{\mathbf{D}_{\lambda_{\normalfont{\text{F}}},\lambda_{\normalfont{\text{Sim}}}}}\|_{2}^{2}$
        \For{$t \geq 1$}
        	\State $\displaystyle
        	    \boldsymbol{\alpha}^{(t)} = \boldsymbol{\gamma}^{(t-1)} - L^{-1}\left(-\nabla \ell\left(\boldsymbol{\gamma}^{(t-1)}\right) + \nabla f_{\mu}\left(\boldsymbol{\gamma}^{(t-1)}\right) \right)$
        	\State $\boldsymbol{\beta}^{(t)} = \mathcal{S}\left(\boldsymbol{\alpha}^{(t)}; L^{-1}\lambda_{\text{Sp} }\right)$
        	\State \textbf{if} $\|\boldsymbol{\beta}^{(t)} - \boldsymbol{\beta}^{(t-1)}\|_{2} \leq \epsilon $ \textbf{break}
        	\State $s^{(t)} = \frac{1 + \sqrt{1 + 4{s^{(t-1)}}^{2}}}{2}$
        	\State $\boldsymbol{\gamma}^{(t)} = \boldsymbol{\beta}^{(t)} + \left(\frac{s^{(t-1)}-1}{s^{(t)}} \right)\left(\boldsymbol{\beta}^{(t)} - \boldsymbol{\beta}^{(t-1)}\right)$
            \State $t \leftarrow t+1$
        \EndFor
        \State \textbf{end for}
        \State $\hat{\boldsymbol{\beta}} \leftarrow \boldsymbol{\beta}^{(t)}$.
      \end{algorithmic}
\end{algorithm}

\section*{Supporting Information}
Web Appendices, Figures and Proofs referenced in Sections 2, 3, 4 and 5 are available with this paper at the Biometrics website on Wiley Online Library. The codes for the proposed JFM is available both on Wiley Online Library and on GitHub: https://github.com/hyungrok-do/joint-fairness-model.

\end{document}